\documentstyle[epsfig]{mn} 
\newif\ifAMStwofonts

\def\nd#1#2{{\rm{d} #1 \over \rm{d} #2}}  
\def\pd#1#2{{\upartial #1 \over \upartial #2}} 
\def\spd#1#2#3{{\upartial ^2 #1 \over \upartial #2 \upartial #3}}
 
\def\hvect#1{{\hat{\bmath{#1}}}} 
\def\det#1{{|{\mathbfss #1}|}} 

\ifoldfss
  \ifCUPmtlplainloaded \else
    \NewTextAlphabet{textbfit} {cmbxti10} {}
    \NewTextAlphabet{textbfss} {cmssbx10} {}
    \NewMathAlphabet{mathbfit} {cmbxti10} {} 
    \NewMathAlphabet{mathbfss} {cmssbx10} {} 
  \fi
  \ifAMStwofonts
    \ifCUPmtlplainloaded \else
      \NewSymbolFont{upmath} {eurm10}
      \NewSymbolFont{AMSa} {msam10}
      \NewMathSymbol{\upi}     {0}{upmath}{19}
      \NewMathSymbol{\umu}     {0}{upmath}{16}
      \NewMathSymbol{\upartial}{0}{upmath}{40}
      \NewMathSymbol{\leqslant}{3}{AMSa}{36}
      \NewMathSymbol{\geqslant}{3}{AMSa}{3E}

    \fi
  \fi
\fi 

\ifnfssone
  \newmathalphabet{\mathit}
  \addtoversion{normal}{\mathit}{cmr}{m}{it}
  \addtoversion{bold}{\mathit}{cmr}{bx}{it}
  \newmathalphabet{\mathbfit} 
  \addtoversion{normal}{\mathbfit}{cmr}{bx}{it}
  \addtoversion{bold}{\mathbfit}{cmr}{bx}{it}
  \newmathalphabet{\mathbfss} 
  \addtoversion{normal}{\mathbfss}{cmss}{bx}{n}
  \addtoversion{bold}{\mathbfss}{cmss}{bx}{n}
  \ifAMStwofonts
    \ifCUPmtlplainloaded \else
      \UseAMStwoboldmath
      \makeatletter
      \new@mathgroup\upmath@group
      \define@mathgroup\mv@normal\upmath@group{eur}{m}{n}
      \define@mathgroup\mv@bold\upmath@group{eur}{b}{n}
      \edef\UPM{\hexnumber\upmath@group}
      \new@mathgroup\amsa@group
      \define@mathgroup\mv@normal\amsa@group{msa}{m}{n}
      \define@mathgroup\mv@bold\amsa@group{msa}{m}{n}
      \edef\AMSa{\hexnumber\amsa@group}
      \makeatother
      \mathchardef\upi="0\UPM19
      \mathchardef\umu="0\UPM16
      \mathchardef\upartial="0\UPM40
      \mathchardef\leqslant="3\AMSa36
      \mathchardef\geqslant="3\AMSa3E
    \fi
  \fi
\fi 

\ifnfsstwo
  \DeclareMathAlphabet{\mathbfit}{OT1}{cmr}{bx}{it}
  \SetMathAlphabet\mathbfit{bold}{OT1}{cmr}{bx}{it}
  \DeclareMathAlphabet{\mathbfss}{OT1}{cmss}{bx}{n}
  \SetMathAlphabet\mathbfss{bold}{OT1}{cmss}{bx}{n}
  \ifAMStwofonts
    \ifCUPmtlplainloaded \else
      \DeclareSymbolFont{UPM}{U}{eur}{m}{n}
      \SetSymbolFont{UPM}{bold}{U}{eur}{b}{n}
      \DeclareSymbolFont{AMSa}{U}{msa}{m}{n}
      \DeclareMathSymbol{\upi}{0}{UPM}{"19}
      \DeclareMathSymbol{\umu}{0}{UPM}{"16}
      \DeclareMathSymbol{\upartial}{0}{UPM}{"40}
      \DeclareMathSymbol{\leqslant}{3}{AMSa}{"36}
      \DeclareMathSymbol{\geqslant}{3}{AMSa}{"3E}
    \fi
  \fi
\fi 

\ifCUPmtlplainloaded \else
  \ifAMStwofonts \else 
    \def\upi{\pi}
    \def\umu{\mu}
    \def\upartial{\partial}
  \fi
\fi

\title[Foreground separation methods] 
{Foreground separation methods for satellite observations of the 
cosmic microwave background} 
\author[M.P. Hobson et al.] 
{M.P.~Hobson$^1$, A.W.~Jones$^1$, A.N.~Lasenby$^1$ and F.R.~Bouchet$^2$ \\
$^{1}$Mullard Radio Astronomy Observatory, Cavendish Laboratory, 
Madingley Road, Cambridge CB3 OHE, UK\\
$^{2}$Institut d'Astrophysique de Paris, CNRS, 98 bis Boulevard Arago, 
F-75014 Paris, France} 
\date{Accepted ???. Received ???; in original form \today}
\pagerange{\pageref{firstpage}--\pageref{lastpage}}
\pubyear{1998} 

\begin{document} 
\maketitle 
\label{firstpage} 

\begin{abstract} 
A maximum entropy method (MEM) is presented for separating the
emission due to different foreground components from simulated satellite
observations of the cosmic microwave background radiation (CMBR). In
particular, the method is applied to simulated observations by the
proposed Planck Surveyor satellite. The simulations, performed by
Bouchet and Gispert (1998), include emission 
from the CMBR, the kinetic and thermal Sunyaev-Zel'dovich (SZ) effects
from galaxy clusters, as well as Galactic dust, free-free and
synchrotron emission. We find that the MEM technique performs well and
produces faithful reconstructions of the main input components.
The method is also compared with traditional Wiener filtering
and is shown to produce consistently better results,
particularly in the recovery of the thermal SZ effect.
\end{abstract} 


\section{Introduction}  
\label{intro} 

The importance of making accurate measurements of the fluctuations
in the CMBR is now widely appreciated. Indeed, by
making maps of these fluctuations and by measuring their power
spectrum, it is hoped that tight constraints may be placed on
fundamental cosmological parameters and that we may 
distinguish between competing theories of structure formation in the
early Universe such as inflation and topological defects.

Several ground-based and balloon-borne experiments are planned over
the next few years, and these should provide accurate images of
the CMBR fluctuations and lead to a significant
improvement in the measurement of the CMBR power spectrum.
Nevertheless, these experiments are unlikely to be able to achieve the
accuracy required to resolve numerous degeneracies that exist in the
parameter set of, for example, the standard inflationary CDM model.
As a result, a new generation of CMBR satellites are 
currently in the final stages of design, and it is hoped that these
experiments will provide definitive measurements of the CMBR power
spectrum as well as detailed all-sky maps of the fluctuations.

According to current estimates, the NASA MAP satellite is due to be launched
in 2000, followed by the ESA Planck Surveyor mission in 2005.
Both experiments aim to make high-resolution, low-noise
maps of the whole sky at several observing frequencies.
As with any CMBR experiment, however, the maps produced will contain
contributions from various foreground components. The main foreground
components are expected to be Galactic dust, free-free and 
synchrotron emission as well as the kinetic and thermal SZ effects 
from galaxy clusters. In addition, significant contamination from 
extragalactic points sources is also likely. 

It is clear that in order to obtain maps of the CMBR fluctuations
alone it is necessary to separate the emission due to these various
components.  The removal of point sources from the satellite
observations is perhaps the most troublesome aspect of this
separation, since our knowledge of the various populations of sources
is incomplete.  Nevertheless, at observing frequencies in the range
10--100 GHz, we expect the point sources to be mainly radio-loud AGN,
including flat-spectrum radiogalaxies and QSOs, blazars and possibly
some inverted-spectrum radiosources. At higher observing frequencies
in the range 300--900 GHz,
the dominant point sources should be infrared luminous
galaxies, radio-quiet AGN and smaller numbers of high-redshift
galaxies and QSOs. However, since the frequency spectra of many of
these extragalactic objects are, in general, rather complicated, any 
extrapolation of their emission over a wide frequency
interval must be performed with caution.

For small fields, a straightforward and effective technique for
removing point sources is to make
high-resolution, high-flux-sensitivity observations of each field, at
frequencies close to those of the CMBR experiment. The point sources
can then be identified and accurately subtracted from the maps
(O'Sullivan et al. 1995).  For multifrequency all-sky satellite
observations, however, such a procedure is infeasible. Nevertheless,
for the Planck Surveyor, it is expected that a significant fraction
of point sources may be identified and removed using the satellite
observations themselves, together perhaps with pre-existing surveys.
Based on the estimated sensitivity of the Planck Surveyor 
to point sources, De Zotti et al. (1997) find 
that it is straightforward, at each observing frequency independently,
to subtract all sources brighter than 1 Jy and that it may be 
possible to subtract all sources brighter than 100 mJy at
intermediate frequencies where the CMBR emission peaks.
Careful modelling of the likely point source contamination
also suggests that the number of pixels affected 
at each frequency should only be a small percentage of the total
number. Moreover, De Zotti et al. find the level of
fluctuations due to unsubtracted sources to be very low. Similar
conclusions follow from the model of Guiderdoni et al. (1997, 1998).
Using simulated maps of point sources (Toffolatti et al. 1998),
a full investigation of their effects on Planck
Surveyor observations will be presented in a forthcoming paper
(Hobson et al, in preparation). 

Aside from extragalactic point sources, the other physical components
mentioned above have reasonably well defined spectral characteristics,
and we may use this information, together with multifrequency
observations, to distinguish between the various foregrounds.  Several
linear methods have been suggested to perform this separation, many of
which are based on Wiener filtering (e.g. Bouchet, Gispert \& Puget
1996; Tegmark \& Efstathiou 1996; Bouchet et al. 1997). 
In this paper, however, we investigate the use of
a non-linear maximum entropy method (MEM) for separating out the
emission due to different physical components and compare its
performance with the Wiener filter approach. We apply these
methods to simulated observations from the Planck surveyor satellite
but, of course, the same algorithms can be used to analyse data from
the MAP satellite. The application of the MEM technique to simulated
interferometer observations of the CMBR is discussed in Maisinger,
Hobson \& Lasenby (1997) and the method has also been applied to the
analysis of ground-based switched-beam observations by Jones et al.
(1997).

\section{Simulated Planck Surveyor observations} 
\label{simobs} 

In order to create simulated Planck Surveyor observations, we must
first build a realistic model of the sky at each of the proposed
observing frequencies. As mentioned above, dust, free-free and
synchrotron emission from our own Galaxy, extragalactic radiosources
and infrared galaxies, and the kinetic and thermal SZ effect from
clusters of galaxies all contribute to sky emission at least at some
frequencies and angular resolutions of interest.  We assume that point
sources may be removed as described above, and that the residual
background of unsubtracted sources is negligible. Thus the simulations
presented here include emission from the three Galactic components, 
the two SZ effects and the primordial CMBR fluctuations.

Simulated maps of these six components are constructed on
$10\times10$-degree fields with 1.5 arcmin pixels; thus each map
consists of $400 \times 400$ pixels. A detailed discussion of these
simulations is given by Bouchet et al. (1997), Gispert \& Bouchet
(1997) and Bouchet \& Gispert (1998).  The primary CMBR fluctuations
are a realisation of a COBE-normalised standard CDM model with
critical density and a Hubble parameter $H_0 = 50$ km s$^{-1}$
Mpc$^{-1}$ (using a program kindly provided by
J.R. Bond). Realisations of the  
kinetic and thermal SZ effects are generated using the Press-Schechter
formalism, as discussed in Aghanim et al. (1997), which yields the number
density of clusters per unit redshift, solid angle and flux density
interval.  The gas profiles of individual clusters are taken as King
models, and their peculiar radial velocities are drawn at random from
an assumed Gaussian velocity distribution with a standard deviation at
$z=0$ of 400 km s$^{-1}$.

For the Galactic dust and free-free emission, 100-$\mu$m IRAS maps are
used as spatial templates. Comparison of dust, free-free and 21cm
{\sc Hi} emission suggests the existence of a spatial correlation
between these components (Kogut et al. 1996; Boulanger et al.
1996). In order to take account of these correlations, the
simulations assume the existence of an {\sc Hi}-correlated component that
accounts for 50 per cent of the free-free emission and 95 per cent of
the dust emission.  The remaining free-free and dust emission is
assumed to come from a second, {\sc Hi}-uncorrelated component. For any
particular simulation, a given 100-$\mu$m IRAS map is used as a
spatial template for the {\sc Hi}-correlated component and a contiguous map
is used for the {\sc Hi}-uncorrelated component.  The dust spectral
behaviour is modelled as a single temperature component at 18 K with
dust emissivity $\propto \nu^2$; the rms level of fluctuations at any
given frequency is scaled accordingly from the 100-$\mu$m IRAS map.
The IRAS map used here has an rms level approximately equal to the
median level for such maps, i.e. about one-half of IRAS 100-$\mu$m
maps of this size have a lower rms, and half have a higher rms when
scales between 1 and 3 degrees are included (see Bouchet et al. 1996
for details). The  
free-free intensity is assumed to vary as $I \propto \nu^{-0.16}$, and
is normalised to give an rms temperature fluctuation of 6.2 $\mu$K at 53 GHz.

No spatial template is available for the synchrotron emission at a
sufficiently high angular resolution, so the simulations of this
component are performed using the 408 MHz radio maps of Haslam et al.
(1982), which have a resolution of 0.85 degrees, and adding to them
(Gaussian) small scale structure that follows a $C_\ell \propto
\ell^{-3}$ power 
spectrum. The synchrotron intensity is assumed vary as $I \propto
\nu^{-0.9}$ and its normalisation taken directly from the 408 MHz maps.
\begin{figure*}
\centerline{\epsfig{
file=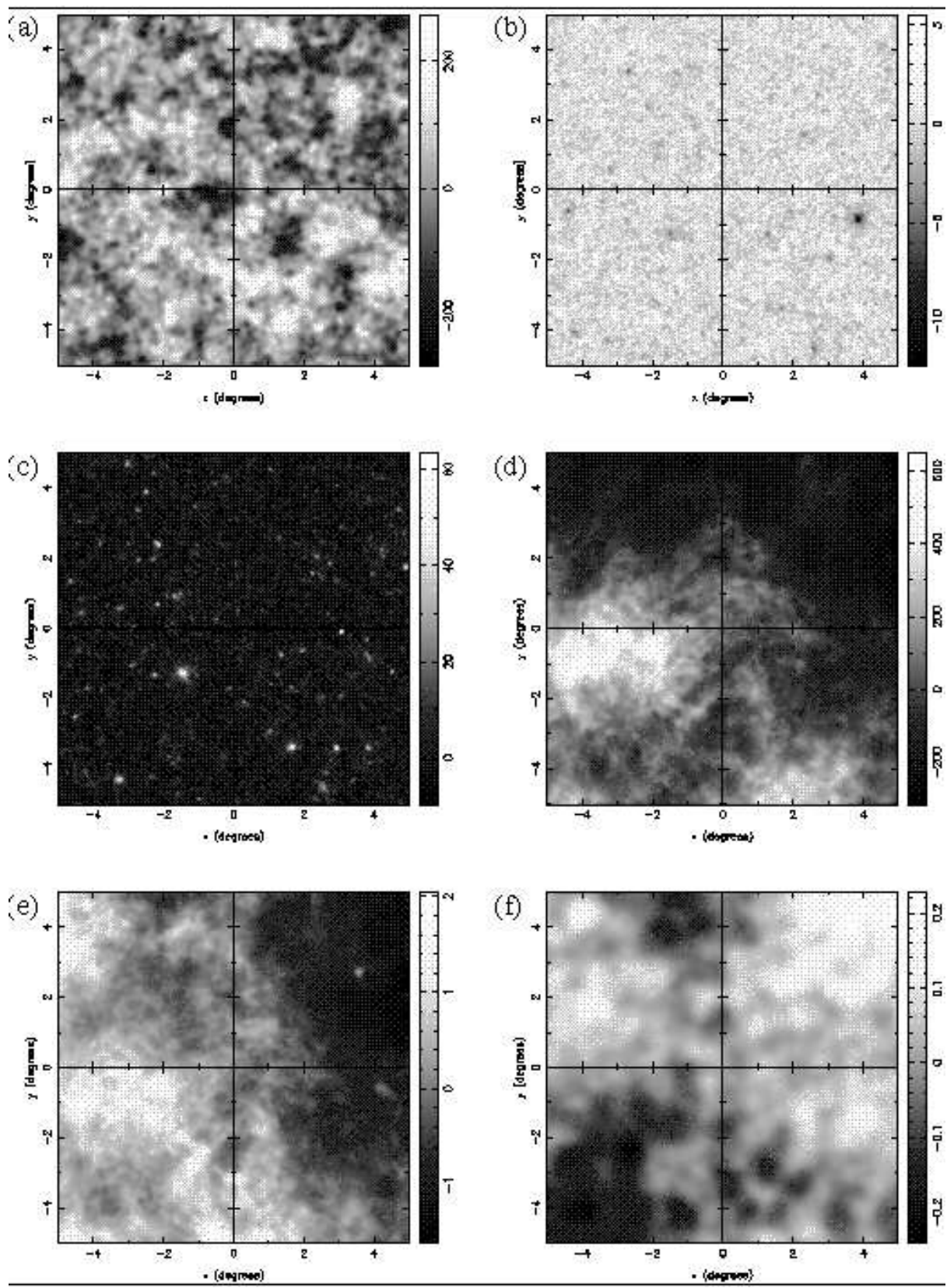,width=16cm}}
\caption{The $10\times 10$-degree realisations of the six input
components used to make simulated Planck Surveyor observations:
(a) primary CMBR fluctuations; (b) kinetic
SZ effect; (c) thermal SZ effect; (d) Galactic dust; (e) Galactic
free-free; (f) Galactic synchrotron emission. 
Each component is plotted at 300 GHz
and has been convolved with a Gaussian beam of FWHM equal to 4.5
arcmin, the maximum angular resolution proposed for the Planck
Surveyor. The map units are equivalent thermodynamic temperature in
$\mu$K.}
\label{fig1}
\end{figure*}

For primary CMBR fluctuations it is usual to work in terms of
temperature rather than intensity. A temperature difference on the sky
$\Delta T_{\rm cmb}(\hvect{x})$ leads to a fluctuation in the
intensity given by
\[
\Delta I_{\rm cmb}(\hvect{x},\nu) 
\approx \left.\pd{B(\nu,T)}{T}\right|_{T=T_0} 
\Delta T_{\rm cmb}(\hvect{x}),
\]
where $B(\nu,T)$ is the Planck function and $T_0=2.726$ K is the
mean temperature of the CMBR (Mather et
al. 1994). The conversion factor can be approximated by
\[
\left.\pd{B(\nu,T)}{T}\right|_{T=T_0}
\approx 24.8\left[\frac{x^2}{\sinh (x/2)}\right]^2
~\mbox{Jy sr$^{-1}$ ($\mu$K)$^{-1}$},
\]
where $x \approx \nu/56.8$ GHz. In order to compare the relative
level of fluctuations in each physical component
we shall adopt the convention of Tegmark \& Efstathiou (1996) and
also define the equivalent thermodynamic temperature fluctuation for
the other components by
\[
\Delta T_p(\hvect{x},\nu) \approx 
\frac{\Delta I_p(\hvect{x},\nu)}{\partial B(\nu,T_0)/\partial T},
\]
where $p$ denotes the relevant physical foreground component. We note that,
in general, the `temperature' fluctuations of these other components
will be frequency dependent, unlike those of the CMBR. For the
remainder of this paper, fluctuations will be quoted in
temperature units measured in $\mu$K.
 
The realisations of the six input components used to make simulated
observations are shown in Fig.~\ref{fig1}.  Each component is plotted
at 300 GHz and, for illustration purposes, has been convolved with a
Gaussian beam of FWHM equal to 4.5 arcmin, which is the highest
angular resolution proposed for the Planck Surveyor.  For
convenience, we have also set the mean of each map to zero, in order
to highlight the relative level of fluctuations due to each component.

From Fig.~\ref{fig1} we see that, as expected, the emission due to
primordial CMBR fluctuations appears Gaussian in nature. This is, of
course, a direct consequence of using a standard inflationary CDM
model to create this realisation. If the CMBR realisation were instead
created assuming an alternative theory of structure formation such as
topological defects, for example, then the CMBR fluctuations are not
required to be Gaussian, but may exhibit sharp edges or highly
non-Gaussian localised hot spots (Bouchet, Bennett \& Stebbins 1988,
Turok 1996).  The emission due to the kinetic and thermal SZ effects
is clearly highly non-Gaussian, being dominated by resolved and
unresolved clusters that appears as sharp peaks of emission. As we
would expect, although an obvious correlation exists between the
positions of the kinetic and thermal SZ effects, the signs and
magnitudes of the kinetic effect are not correlated with those of the
thermal effect.  We also note that the IRAS 100-$\mu$m maps used as
templates for the Galactic dust and free-free emission also appear
quite non-Gaussian; the imposed correlation between the dust and
free-free emission is also clearly seen. Finally, the synchrotron
emission seems quite Gaussian, although this appearance is due mainly
to the addition to the Haslam 408 MHz map of Gaussian small scale
structure, following a $C_\ell \propto \ell^{-3}$ power law, on
angular scales below 0.85 degrees.

\begin{figure}
\centerline{\epsfig{
file=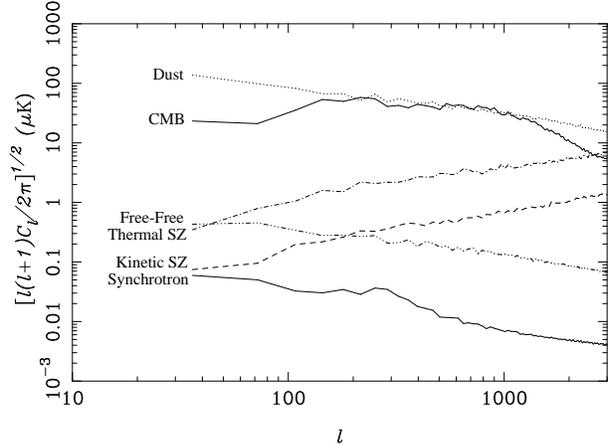,width=8cm}}
\caption{The azimuthally-averaged power spectra of the input maps 
at 300~GHz.}
\label{fig2}
\end{figure}

The azimuthally-averaged power spectra of the input maps are shown in
Fig.~\ref{fig2}.  At lower multipoles, all three Galactic components
have power spectra which vary roughly as $C_\ell \propto \ell^{-3}$
(for the synchrotron component small scale structure with this power
spectrum was added artificial for $\ell \ga 250$).  
For the kinetic and thermal SZ effects, however, the power spectra are quite
different and are better approximated by a white-noise power spectrum
$C_\ell \propto {\rm constant}$, as expected for Poisson-distributed
processes.

Using the realisations for each physical component shown in
Fig.~\ref{fig1}, it is straightforward to simulate Planck Surveyor
observations. The satellite is made up of two mains parts: the Low
Frequency Instrument (LFI), which uses HEMT radio receivers, and the
High Frequency Instrument (HFI), which contains bolometer arrays.
Since the final design of the satellite is still undecided, the
precise values of observational parameters for the LFI and HFI are
subject to revision.  Nevertheless, recent proposed changes to both
instruments may significantly improve the sensitivity of the
satellite, as compared to the design outlined in the ESA phase A study
(Bersanelli et al. 1996). Therefore, although these modifications are
not yet finalised, we have incorporated the latest design
specifications into our simulations. The parameters used in making the
simulated observations are given in Table~\ref{table1}.
%
\begin{table*}
\begin{center}
\caption{Proposed observational parameters for the Planck Surveyor
satellite (Efstathiou, private communication). 
Angular resolution is quoted as FWHM 
for a Gaussian beam. Sensitivities are quoted per
FWHM for 12 months of observation.}
\begin{tabular}{lccccccccccc} \hline
& \multicolumn{4}{c}{Low Frequency Instrument} 
& & \multicolumn{6}{c}{High Frequency Instrument} \\ \hline
Central frequency (GHz):   
& 30   & 44   & 70   & 100  & & 100  & 143  & 217  & 353  & 545  & 857 \\
Fractional bandwidth ($\Delta\nu/\nu$):   
& 0.2 & 0.2 & 0.2 & 0.2 & & 0.37 & 0.37 & 0.37 & 0.37 & 0.37 & 0.37 \\
Transmission:
& 1.0  & 1.0  & 1.0  & 1.0  & & 0.3  & 0.3  & 0.3  & 0.3  & 0.3  & 0.3 \\
Angular resolution (arcmin):  
& 33   & 23   & 14   & 10   & & 10.6   & 7.4  & 4.9  & 4.5  & 4.5  & 4.5 \\
$\Delta T$ sensitivity ($\mu$K): 
& 4.4  & 6.5  & 9.8 & 11.7 & & 4.9 & 5.7 & 12.5  & 40.9 & 392  & 12621 \\
\hline
\end{tabular}
\end{center}
\label{table1}
\end{table*}

The simulated observations are produced by integrating the emission
due to each physical component across each waveband, assuming the
transmission is uniform across the band. At each observing frequency,
the total sky emission is convolved with a Gaussian beam of the
appropriate FWHM. Finally, isotropic noise is added to the maps,
assuming a spatial sampling rate of FWHM/2.4 at each frequency (thus
the noise rms of the maps is about 2.4 times {\em higher} than the
instrumental sensitivity per FWHM quoted in Table~\ref{table1}). We
note, however, that the assumption of isotropic noise is not required by the
separation algorithms discussed in Section \ref{methods}. We have also
assumed that any striping due to the scanning strategy and $1/f$ noise 
has been removed to sufficient accuracy that any residuals are negligible.

\begin{figure}
\centerline{\epsfig{
file=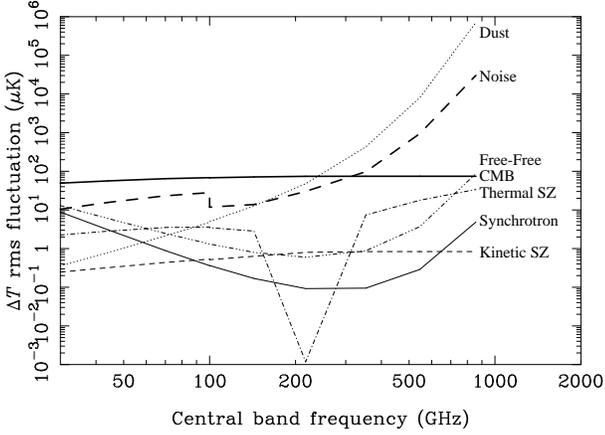,width=8cm}}
\caption{The rms thermodynamic temperature fluctuations at each
Planck Surveyor 
observing frequency due to each physical component, after convolution
with the appropriate beam and using a sampling rate of FWHM/2.4. The
rms noise per pixel at each frequency channel is also plotted.}
\label{fig3}
\end{figure}

Fig.~\ref{fig3} shows the rms temperature fluctuations at each
observing frequency due to each physical component, after convolution
with the appropriate beam.  The rms noise per pixel at each frequency
channel is also plotted.
We see from the figure that, as expected, the rms temperature
fluctuation of the CMBR is almost constant across the frequency
channels; the only variation being due to the convolution with
beams of different sizes. Furthermore, for all channels up to 217 GHz, the
CMBR signal is several times the level of the instrumental noise even
for a pixelisation at FWHM/2.4 (which boosts by 2.4 the noise level
per FWHM). At
higher frequencies, the noise level exceeds the CMBR signal but is
itself dominated by Galactic dust emission.  We also see a sharp dip in
the rms level of the thermal SZ effect at 217 GHz, 
since the emission from this component is close to
zero at this frequency. At any given frequency, the rms level of the
thermal SZ effect is at least an order of magnitude below that of the
dominant component. The kinetic SZ effect has the
same spectral characteristics as the CMBR, but the effect of
convolution with beams of different sizes has a significant effect on the
point-like emission and leads to a more pronounced variation in the
observed rms level than for the CMBR (since then most of the power is
at small scales). The observed rms level 
of the kinetic SZ is at
least two orders of magnitude below the dominant component at any
given frequency. In a similar manner, the Galactic free-free and
synchrotron emission are also completely dominated by either CMBR or
dust emission at all observing frequencies.

\begin{figure*}
\centerline{\epsfig{
file=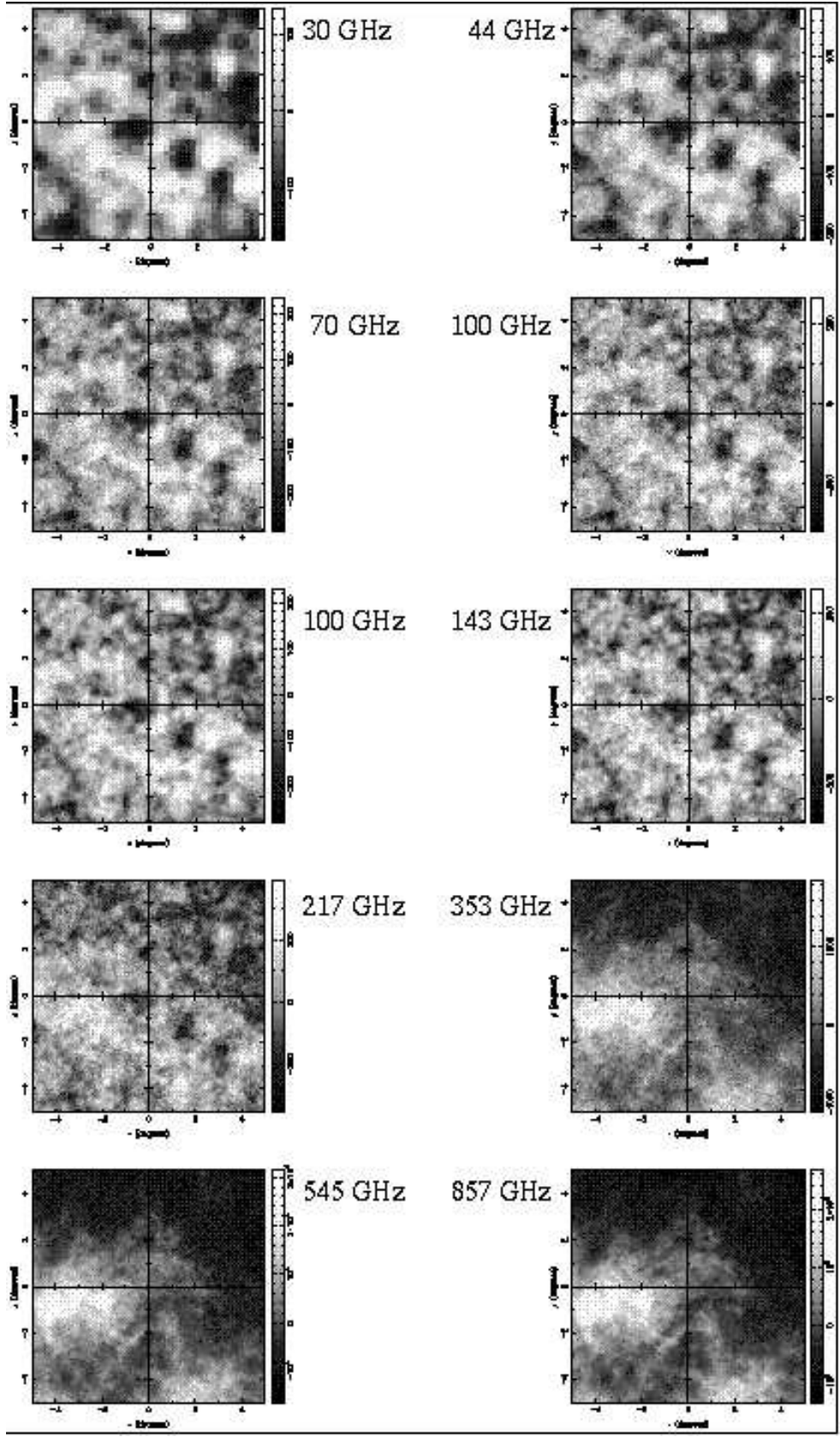,width=12.95cm}}
\caption{The $10\times 10$-degree maps observed at each of the ten
Planck Surveyor frequencies listed in Table~\ref{table1}.
At each frequency we assume a Gaussian beam with the appropriate FWHM and
a sampling rate of FWHM/2.4. Isotropic noise with the relevant rms 
has been added to each map. The map units are 
equivalent thermodynamic temperature in $\mu$K.}
\label{fig4}
\end{figure*}

The observed maps at each of the ten Planck Surveyor 
frequencies are shown in
Fig.~\ref{fig4} in units of equivalent thermodynamic temperature
measured in $\mu$K.
The coarser pixelisation at the lower observing frequencies is due to the
FWHM/2.4 sampling rate. Moreover, at these lower frequencies,
the effect of convolution with the relatively large beam is also 
easily seen. As the observing frequency increases, the beam size
becomes smaller, leading to a corresponding increase in the sampling
rate. Consequently, the observed maps more closely resemble
the input map of the dominant physical component at each frequency.
As may have been anticipated from Fig.~\ref{fig3}, the emission
in the lowest seven channels is dominated by the CMBR, whereas dust
emission dominates in the highest three channels. Indeed, the main
reason for the inclusion of the highest frequency channels is
to obtain an accurate dust model, in order that it may be subtracted
from lower frequency channels with some confidence. Perhaps the most
notable feature of the ten channels maps is that, at least by eye, it
is not possible to discern features due to physical components other
than the CMBR or dust. 

\section{Component separation methods}
\label{methods}

As a first step in discussing component separation methods, 
let us consider in more detail how the
simulated data are made. At any given frequency $\nu$ 
the total rms temperature fluctuation on the sky in a direction
$\hvect{x}$ is given by the superposition of $n_c$ physical
components ($n_c=6$ in our simulations). It is convenient to factorise
the contribution of each process into a spatial template $s_p(\hvect{x})$
at a reference frequency $\nu_0$ and a frequency dependence
$f_p(\nu)$, so that
\[
\Delta T(\hvect{x},\nu) = \sum_{p=1}^{n_c} \Delta T_p(\hvect{x},\nu) =
\sum_{p=1}^{n_c} f_p(\nu)s_p(\hvect{x}).
\]
In this paper we take the reference frequency $\nu_0 = 300$ GHz 
and normalise the frequency dependencies such that
$f_p(\nu_0)=1$.

If we observe the sky at $n_f$ observing frequencies then, in any given
direction $\hvect{x}$, we obtain a $n_f$-component {\em data vector}
that contains the observed temperature fluctuation in this direction
at each observing frequency plus instrumental noise.
In order to relate this data vector to the
emission from each physical component it is useful 
to introduce the $n_f\times n_c$ {\em frequency response matrix}
with components defined by
\begin{equation}
F_{\nu p} = \int_0^\infty t_\nu(\nu') f_p(\nu') \,{\rm d}\nu'
\label{freqresp}
\end{equation}
where $t_\nu(\nu')$ is the frequency response (or transmission) of the
$\nu$th frequency channel. Assuming that the satellite observing beam in each
channel is spatially invariant, we may write the beam-smoothing
as a convolution and, in discretised form, the $\nu$th component of
the data vector in the direction $\hvect{x}$ is then given by
\begin{equation}
d_\nu(\hvect{x}) = \sum_{j=1}^{N_p} P_\nu(|\hvect{x}-\hvect{x}_j|) 
\sum_{p=1}^{n_c} F_{\nu p}\,s_p(\hvect{x}_j) + \epsilon_\nu(\hvect{x})
\label{datadef}
\end{equation}
where $P_\nu(\hvect{x})$ is the beam profile for the $\nu$th frequency
channel, and the index $j$ labels the $N_p$ pixels in each of the
simulated input maps shown in Fig.~\ref{fig1}; the
$\epsilon_\nu(\hvect{x})$ term represents the instrumental noise in
the $\nu$th channel in the direction $\hvect{x}$. 

In each channel the beam profile is assumed spatially invariant and the noise
statistically homogeneous (which are both reasonable assumptions for
small fields), and it is more convenient to work in Fourier
space, since the convolution in (\ref{datadef}) becomes a simple
multiplication and we obtain
\begin{equation}
\widetilde{d}_\nu(\bmath{k}) 
= \sum_{p=1}^{n_c} R_{\nu p}(\bmath{k})\widetilde{s}_p(\bmath{k})
+\widetilde{\epsilon}_\nu(\bmath{k}),
\label{dataft}
\end{equation}
where $R_{\nu p}(\bmath{k}) = \widetilde{P}_\nu(\bmath{k})F_{\nu p}$ are
the components of the {\em response matrix} for the observations.
It is important to
note that (\ref{dataft}) is satisfied at each 
Fourier mode $\bmath{k}$ {\em independently}.
Thus, in matrix notation, at each mode we have
\begin{equation}
\mathbfss{d} = \mathbfss{R}\mathbfss{s}+\bmath{\epsilon}
\label{dataft2}
\end{equation}
where $\mathbfss{d}$, $\mathbfss{s}$ and $\bmath{\epsilon}$ are column vectors
containing $n_f$, $n_c$ and $n_f$ complex components respectively, and
the response matrix $\mathbfss{R}$ has dimensions $n_f\times
n_c$. Although the column vectors in (\ref{dataft2}) refer to
quantities defined in the Fourier domain, it should be noted that
for later convenience they are not written with a tilde. 

The significant simplification that results from working in the
Fourier domain is clear, since the dimensions of the matrices in
(\ref{dataft2}) are rather small ($n_c=6$ and $n_f=10$ in our
simulations). Thus, the situation reduces to the solving a small-scale
linear inversion problem at each Fourier mode {\em separately}. Once
this inversion has been performed for all the measured modes, the
spatial templates for the sky emission due to each physical component
at the reference frequency $\nu_0$ are then obtained by an inverse
Fourier transformation. Owing to the presence of instrumental noise, however,
it is clear that the inverse,
${\mathbfss R}^{-1}$, of the response matrix at each Fourier mode does
not exist and that the linear inversion problem in each case is
degenerate. The approximate inversion of (\ref{dataft2}) must therefore
be performed using a statistical technique in which the inversion is
regularised in some way.  This naturally leads us to consider a
Bayesian approach.

\subsection{Bayes' theorem}
\label{bayest}

Bayes' theorem states that, given a hypothesis $H$ and some data $D$
the posterior probability $\Pr(H|D)$ is the product of the likelihood
$\Pr(D|H)$ and the prior probability $\Pr(H)$, normalised by the
evidence $\Pr(D)$,
\[ 
\Pr (H|D) = \frac{\Pr(H)\Pr(D|H)}{\Pr(D)}.
\]

In our application, we consider each Fourier mode
$\bmath{k}$ separately and, from (\ref{dataft2}), we see that the data
consist of the $n_f$ complex numbers in the data vector 
$\mathbfss{d}$, and we take the `hypothesis' to consist of the
$n_c$ complex numbers in the signal vector $\mathbfss{s}$. 
We then choose as our estimator $\hat{\mathbfss{s}}$ of the signal vector 
that which 
maximises the posterior probability $\Pr(\mathbfss{s}|\mathbfss{d})$. Since
the evidence in Bayes' theorem is merely a normalisation constant
we must therefore maximise with respect to $\mathbfss{s}$ the
quantity
\begin{equation}
\Pr(\mathbfss{s}|\mathbfss{d}) \propto
\Pr(\mathbfss{d}|\mathbfss{s}) \Pr(\mathbfss{s})
\label{bayes}
\end{equation}
which is the product of the likelihood
$\Pr(\mathbfss{d}|\mathbfss{s})$ and the prior $\Pr(\mathbfss{s})$.

Let us first consider the form of the likelihood.
If the instrumental noise on each frequency channel is
Gaussian-distributed, then at each Fourier mode the probability 
distribution of the $n_f$-component noise vector
$\bmath{\epsilon}$ is described by an
$n_f$-dimensional multivariate Gaussian. Assuming the
expectation value of the noise to be zero at each observing frequency,
the likelihood is therefore given by
\begin{eqnarray}
\Pr ({\mathbfss d}|{\mathbfss s}) 
& \propto & \exp \left(-\bmath{\epsilon}^\dagger
{\mathbfss N}^{-1}\bmath{\epsilon}\right) \nonumber \\
& \propto & \exp 
\left[-({\mathbfss d}-{\mathbfss R}{\mathbfss s})^\dagger 
{\mathbfss N}^{-1} ({\mathbfss d}-{\mathbfss R}{\mathbfss s})\right],
\label{likedef}
\end{eqnarray}
where the dagger denotes the Hermitian conjugate and in the second line
we have used (\ref{dataft2}). We note that no factor of $1/2$ appears in the
exponent in (\ref{likedef}) since it refers to the multivariate Gaussian
distribution of a set of complex random variables.
The {\em noise covariance matrix} $\mathbfss{N}$ 
has dimensions $n_f \times n_f$ and at any given Fourier mode $\bmath{k}$
it is defined by 
\begin{equation}
{\mathbfss N}(\bmath{k}) = \langle 
\bmath{\epsilon}(\bmath{k})
\bmath{\epsilon}^\dagger (\bmath{k})
\rangle, 
\label{ncovdef}
\end{equation}
i.e. its elements are given by
$N_{\nu\nu'}(\bmath{k}) = \langle\,\widetilde{\epsilon}_\nu(\bmath{k})\,
\widetilde{\epsilon}_{\nu'}^{\,*}(\bmath{k})\rangle$, 
where the asterisk denotes complex conjugation. 
Thus, at a given
Fourier mode, the $\nu$th diagonal element of $\mathbfss{N}$ contains the
value at that mode of the ensemble-averaged power spectra of the
instrumental noise on the $\nu$th frequency channel. If the noise is
uncorrelated between channels then the off-diagonal elements are 
zero for all $\bmath{k}$. 

We note that the expression in square brackets in (\ref{likedef})
is simply the $\chi^2$ misfit statistic. Since, for a given set of
observations, the data vector $\mathbfss{d}$, the response
matrix $\mathbfss{R}$ and the noise covariance matrix $\mathbfss{N}$ 
are all fixed, we may consider the misfit statistic as a function only of
the signal vector $\mathbfss{s}$, 
\begin{equation}
\chi^2({\mathbfss s}) = ({\mathbfss d}-{\mathbfss R}{\mathbfss s})^\dagger 
{\mathbfss N}^{-1} ({\mathbfss d}-{\mathbfss R}{\mathbfss s}),
\label{chi2def}
\end{equation}
so that the likelihood can be written as 
\begin{equation}
\Pr({\mathbfss d}|{\mathbfss
s}) \propto \exp[-\chi^2({\mathbfss s})].
\label{likehd}
\end{equation}
Having calculated the form of the likelihood
we must now turn our attention to the form of the prior probability
$\Pr(\mathbfss{s})$.

\subsection{The Gaussian prior}
\label{gaussprior}

If we assume that the emission due to each of the physical components
shown in Fig.~\ref{fig1}
is well approximated by a Gaussian random field, then it is straightforward
to derive an appropriate form for the prior $\Pr(\mathbfss{s})$. 
In this case, the probability distribution of the sky emission
is described by a multivariate Gaussian distribution, 
characterised by a given sky covariance matrix. Thus, at each mode
$\bmath{k}$ in Fourier space, the probability distribution of the
signal vector $\mathbfss{s}$ is also described by a multivariate
Gaussian of dimension $n_c$, where $n_c$ is the number of distinct physical
components ($n_c=6$ in our simulations). The prior therefore has the form
\begin{equation}
\Pr({\mathbfss s}) \propto 
\exp \left(-{\mathbfss s}^\dagger{\mathbfss C}^{-1}{\mathbfss s}\right),
\label{gpdef}
\end{equation}
where the signal covariance matrix $\mathbfss{C}$ is real with dimensions
$n_c \times n_c$ and is given by
\begin{equation}
{\mathbfss C}(\bmath{k}) = \langle 
{\mathbfss s}(\bmath{k})
{\mathbfss s}^\dagger (\bmath{k})
\rangle,
\label{covdef}
\end{equation}
i.e. it has elements
$C_{pp'}(\bmath{k}) = \langle\, \widetilde{s}_p(\bmath{k})\,
\widetilde{s}_{p'}^{\,*}(\bmath{k})\rangle$.
Thus, at each Fourier mode, the $p$th diagonal element of 
$\mathbfss{C}$ contains the
value of the ensemble-averaged power spectrum of the $p$th 
physical component at the reference frequency $\nu_0$;
the off-diagonal terms describe cross-power spectra between the components.

Strictly speaking, the use of this prior requires advance knowledge of the
full covariance structure of the processes that we are trying to
reconstruct. Nevertheless, it is anticipated that some information
concerning the power spectra of the various components, and
correlations between them, will be available either from pre-existing
observations or by performing an initial approximate separation using,
for example, the singular value decomposition (SVD) algorithm 
(see Bouchet et al.
1997, Bouchet \& Gispert 1998). A discussion of the SVD solution, in
the context of Bayes' theorem, is given in Appendix C.
This information can then be used to
construct an approximate 
signal covariance matrix for use in $\Pr(\mathbfss{s})$.

Substituting (\ref{likehd}) and (\ref{gpdef}) into (\ref{bayes}), 
the posterior probability is then given by
\begin{equation}
\Pr({\mathbfss s}|{\mathbfss d}) \propto 
\exp 
\left[-\chi^2({\mathbfss s})
-{\mathbfss s}^\dagger{\mathbfss C}^{-1}{\mathbfss s}\right].
\label{gausspost}
\end{equation}
where $\chi^2({\mathbfss s})$ is given by (\ref{chi2def}). 
Completing the square for $\mathbfss{s}$ in the exponential
(see Zaroubi et al. 1995), it is
straightforward to show that the posterior probability is also a
multivariate Gaussian of the form
\begin{equation}
\Pr({\mathbfss s}|{\mathbfss d}) 
\propto
\exp 
\left[-({\mathbfss s}-\hat{{\mathbfss s}})^\dagger 
{\mathbfss E}^{-1} ({\mathbfss s}-\hat{{\mathbfss s}})\right].
\label{gpost}
\end{equation}
which has its maximum value at the estimate
$\hat{\mathbfss{s}}$ of the signal vector and where $\mathbfss{E}$ is the 
covariance matrix of the reconstruction errors.

The estimate $\hat{\mathbfss{s}}$ of the signal vector is found to be
\begin{equation}
\hat{{\mathbfss s}} = 
\left({\mathbfss C}^{-1}
+{\mathbfss R}^\dagger{\mathbfss N}^{-1}{\mathbfss R}\right)^{-1}
{\mathbfss R}^\dagger{\mathbfss N}^{-1}{\mathbfss d} 
\equiv {\mathbfss W}{\mathbfss d},
\label{wfrecon}
\end{equation}
where we have identified the Wiener matrix $\mathbfss{W}$. Thus, we
find that by
assuming a Gaussian prior of the form (\ref{gpdef}) in Bayes' theorem,
we recover the standard Wiener filter. This optimal linear filter is
usually derived by choosing the elements of $\mathbfss{W}$ such that
they minimise the variances of the resulting reconstruction
errors. From (\ref{wfrecon}) we see that
at a given Fourier mode, we may calculate the 
estimator $\hat{\mathbfss{s}}$ that maximises the posterior probability
simply by multiplying the data vector $\mathbfss{d}$ by the Wiener
matrix $\mathbfss{W}$. Equation (\ref{wfrecon}) can also be derived 
straightforwardly by differentiating (\ref{gausspost}) with respect
$\mathbfss{s}$ and equating the result to zero (see Appendix A).

As is well-known, the assignment
of errors on the Wiener filter reconstruction is straightforward and the 
covariance matrix of the reconstruction errors $\mathbfss{E}$ 
in (\ref{wfrecon}) is given by
\begin{equation}
{\mathbfss E} 
\equiv \langle 
({\mathbfss s}-\hat{\mathbfss s})
({\mathbfss s}-\hat{\mathbfss s})^\dagger
\rangle
= \left({\mathbfss C}^{-1}
+{\mathbfss R}^\dagger{\mathbfss N}^{-1}{\mathbfss R}\right)^{-1}
\label{wferrors}
\end{equation}
Since the posterior probability (\ref{gpost}) is Gaussian, 
this matrix is simply the inverse Hessian or curvature matrix
of (minus) the exponent in (\ref{gpost}), evaluated at 
$\hat{\mathbfss s}$ (see Appendix A).

It should be noted that the linear nature of the Wiener
filter and the simple propagation of errors are both direct consequences of
assuming that the spatial templates we wish to reconstruct are
well-described by Gaussian random fields with a known covariance
structure. Several applications are given in Bouchet et al. (1997).

\subsection{The entropic prior}
\label{entprior}

It is clear from Fig.~\ref{fig1} that the emission due to several of the
underlying physical processes is far from Gaussian. This is
particularly pronounced for the kinetic and thermal SZ effects, but
the Galactic dust and free-free emissions also appear quite
non-Gaussian. Ideally, one might like to assign priors for the various
physical components by measuring empirically the probability distribution of
temperature fluctuations from numerous realisations of each component.
This is not feasible in practice, however, and instead we consider
here the use of the entropic prior, which is based on
information-theoretic considerations alone. 

Let us consider a discretised image $h_j$ consisting of $L$ cells, so
that $j=1,\ldots, L$; we may consider the $h_j$ as
the components of an image vector $\mathbfss{h}$.
Using very general notions of
subset independence, coordinate invariance and system independence, it
may be shown (Skilling 1989) that the prior probability assigned to
the values of 
the components in this vector should take form
\[
\Pr({\mathbfss h}) \propto \exp[\alpha S({\mathbfss h},{\mathbfss m})],
\]
where the dimensional constant $\alpha$ depends on the scaling
of the problem and may be considered as a regularising parameter, and
$\mathbfss{m}$ is a model vector to which $\mathbfss{h}$ defaults in
the absence of any data.
The function $S(\mathbfss{h},\mathbfss{m})$ is the {\em cross entropy} of 
$\mathbfss{h}$ and $\mathbfss{m}$. In standard
applications of the maximum entropy method, the image $\mathbfss{h}$ 
is taken to be a positive additive distribution (PAD). Nevertheless, the MEM
approach can be extended to images that take both positive and
negative values by considering them to be the difference of two PADS, so that
\[
\mathbfss{h}= \mathbfss{u}-\mathbfss{v}.
\]
where $\mathbfss{u}$ and $\mathbfss{v}$ are the positive and
negative parts of $\mathbfss{h}$ respectively.
In this case, the cross entropy is given by (Gull \& Skilling 1990; 
Hobson \& Lasenby 1998)

\begin{equation}
S({\mathbfss h},{\mathbfss m}_u,{\mathbfss m}_v) =  
\sum_{j=1}^{L}\left\{
\psi_j -{m_u}_j-{m_v}_j
-h_j\ln\left[\frac{\psi_j+h_j}{2{m_u}_j}\right]\right\}, 
\label{entdef}
\end{equation}
where $\psi_j = [h_j^2+4{m_u}_j{m_v}_j]^{1/2}$ and ${\mathbfss m}_u$ and
${\mathbfss m}_v$ are separate models for each PAD. The global maximum of the
cross entropy occurs at ${\mathbfss h} = {\mathbfss m}_u-{\mathbfss m}_v$.

In our application, we might initially suppose that at each Fourier
mode we should take the `image' to be the $n_c$ components of the
signal vector $\mathbfss{s}$.  However, this results in two
additional complications. First, the components of signal vector are,
in general, complex, but the cross entropy given in (\ref{entdef}) is
defined only if the image $\mathbfss{h}$ is real.  Nevertheless, the
MEM technique can be straightforwardly extended to the reconstruction of
a complex image by making a slight modification to the above
discussion. If the image $\mathbfss{h}$ is complex, then models 
${\mathbfss m}_u$ and ${\mathbfss m}_v$ are also taken to be complex.
In this case, the real and imaginary parts of ${\mathbfss m}_u$ are
the models for the positive portions of the real and
imaginary parts of $\mathbfss{h}$ respectively.
Similarly, the real and imaginary
parts of ${\mathbfss m}_v$ are the models for the negative portions of
the real and imaginary parts of the image. 
The total cross entropy is then obtained by
evaluating the sum (\ref{entdef}) using first the real parts and then
the imaginary parts of ${\mathbfss h}$, ${\mathbfss m}_u$ and
${\mathbfss m}_v$, and adding the results. Thus the total cross
entropy for the complex image $\mathbfss{h}$ is given by
%
\begin{equation}
S(\Re({\mathbfss h}),\Re({\mathbfss m}_u),\Re({\mathbfss m}_v)) +
S(\Im({\mathbfss h}),\Im({\mathbfss m}_u),\Im({\mathbfss m}_v)),
\label{totent}
\end{equation}
where $\Re$ and $\Im$ denote the
real and imaginary parts of each vector. For simplicity we denote the
sum (\ref{totent}) by $S_c({\mathbfss h},{\mathbfss m}_u,{\mathbfss
m}_v)$ where the subscript $c$ indicates that it is the entropy of a
complex image.

The second complication mentioned above
is more subtle and results from the fact that
one of the fundamental axioms of the MEM is that it should not itself
introduce correlations between individual elements of the image.
However, as discussed in previous subsection, the elements
of the signal vector $\mathbfss{s}$ at each Fourier mode may well be
correlated, this correlation being described by the signal covariance
matrix $\mathbfss{C}$ defined in (\ref{covdef}). Moreover, if prior
information is available concerning these correlations, we would wish
to include it in our analysis. We are therefore lead
to consider the introduction of an intrinsic correlation function
(ICF) into the MEM framework (Gull \& Skilling 1990).

The inclusion of an ICF is most easily achieved by assuming that, at
each Fourier mode, the `image' does not consist of the components of
the signal vector $\mathbfss{s}$, but that instead $\mathbfss{h}$
consists of the components of a vector of hidden variables that are
related to the signal vector by
\begin{equation}
\mathbfss{s}=\mathbfss{L}\mathbfss{h},
\label{icfdef}
\end{equation}
The $n_c \times n_c$ lower triangular matrix $\mathbfss{L}$ in
(\ref{icfdef}) is that obtained by performing a Cholesky decomposition
of the signal covariance matrix, i.e.
$\mathbfss{C}=\mathbfss{L}\mathbfss{L}^{\rm T}$. We note that since
$\mathbfss{C}$ is real then so is $\mathbfss{L}$.
Thus, if the
components of $\mathbfss{h}$ are apriori uncorrelated 
(thereby satisfying the MEM axiom) and of unit variance,
so that $\langle\,h_p\,h_{p'}^{\,*}\,\rangle = \delta_{pp'}$, we
find that, as required, the {\em a priori} 
covariance structure of the signal vector
is given by
\[
\langle\mathbfss{s}\mathbfss{s}^\dagger\rangle
=\langle\mathbfss{L}\mathbfss{h}\mathbfss{h}^\dagger
\mathbfss{L}^{\rm T}\rangle
=\mathbfss{L}\langle\mathbfss{h}\mathbfss{h}^\dagger\rangle
\mathbfss{L}^{\rm T} = \mathbfss{L}\mathbfss{L}^{\rm T} = \mathbfss{C}.
\]
Moreover, using this construction the
expected rms level for the real or imaginary part of each element of
$\mathbfss{h}$ is simply equal
to $1/\sqrt{2}$. Therefore, at each Fourier mode, we assign the 
real and imaginary parts of every component in the model vectors 
${\mathbfss m}_u$ and ${\mathbfss m}_v$ to be equal to $m=1/\sqrt{2}$.

Substituting (\ref{icfdef}) into (\ref{chi2def}), $\chi^2$ can also be
written in terms of $\mathbfss{h}$ and is given by
\begin{equation}
\chi^2({\mathbfss h}) = 
({\mathbfss d}-{\mathbfss R}{\mathbfss L}{\mathbfss h})^\dagger 
{\mathbfss N}^{-1} ({\mathbfss d}-{\mathbfss R}{\mathbfss L}{\mathbfss h}).
\label{chi2def2}
\end{equation}
Thus, using an entropic prior, the posterior probability becomes
\begin{equation}
\Pr({\mathbfss h}|{\mathbfss d})
\propto \exp\left[-\chi^2({\mathbfss h})
+\alpha S_c({\mathbfss h},{\mathbfss m}_u,{\mathbfss m}_v)\right].
\label{postmem}
\end{equation}
where the cross entropy $S_c({\mathbfss h},{\mathbfss m}_u,{\mathbfss
m}_v)$ is given by (\ref{totent}) and (\ref{entdef}). 

\subsection{Maximising the posterior probability}
\label{maxpost}

As discussed in Section \ref{bayest}, we choose our estimate
$\hat{\mathbfss{s}}$ of the signal vector at each Fourier mode, as
that which maximises the posterior probability $\Pr({\mathbfss
s}|{\mathbfss d})$ with respect to $\mathbfss{s}$.

For the Gaussian prior, we found in subsection \ref{gaussprior} that
the posterior probability is also a Gaussian and that the estimate
$\hat{\mathbfss{s}}$ is given directly by the linear relation 
(\ref{wfrecon}). Nevertheless, 
we also note that, in terms of $\mathbfss{h}$ defined in
(\ref{icfdef}), the quadratic form 
in the exponent of the Gaussian prior (\ref{gpdef}) has the
particularly simple form
\[
{\mathbfss s}^\dagger{\mathbfss{C}^{-1}}{\mathbfss s}
= {\mathbfss h}^\dagger{\mathbfss L}^{\rm T}
({\mathbfss L}{\mathbfss L}^{\rm T})^{-1}{\mathbfss L}{\mathbfss h}
= {\mathbfss h}^\dagger{\mathbfss L}^{\rm T}
({\mathbfss L}^{\rm T})^{-1}{\mathbfss L}^{-1}{\mathbfss L}{\mathbfss h}
={\mathbfss h}^\dagger{\mathbfss h},
\]
i.e. it is equal to the inner product of $\mathbfss{h}$ with itself.
Thus, using a Gaussian prior, the posterior probability can be written
in terms of $\mathbfss{h}$ as
\begin{equation}
\Pr({\mathbfss h}|{\mathbfss d})
\propto \exp\left[-\chi^2({\mathbfss h})
-{\mathbfss h}^\dagger{\mathbfss h}\right].
\label{postgh}
\end{equation}
where $\chi^2({\mathbfss h})$ is given by (\ref{chi2def2})
Therefore, in addition to using the linear relation (\ref{wfrecon}),
the Wiener filter estimate $\hat{\mathbfss{s}}$ can also
be found by first minimising the function
\begin{equation}
\Phi_{\rm WF}({\mathbfss h}) 
= \chi^2({\mathbfss h})+{\mathbfss h}^\dagger{\mathbfss h},
\label{wffunc}
\end{equation}
to obtain the estimate $\hat{\mathbfss{h}}$ of the corresponding
hidden vector, and then using (\ref{icfdef}) to give
$\hat{{\mathbfss s}} = {\mathbfss L}\hat{{\mathbfss h}}$.


We have developed an algorithm (which will be presented in a forthcoming
paper) for minimising the function $\Phi_{\rm WF}$ with respect
to ${\mathbfss h}$. Indeed, this algorithm calculates the reconstruction
$\hat{\mathbfss h}$ in slightly less
CPU time than the matrix inversions and multiplications required to
evaluate the linear relation (\ref{wfrecon}). The minimiser requires
only the first derivatives of the function and these are given in
Appendix A.

Let us now consider the MEM solution. From (\ref{postmem}), we see
that maximising the posterior probability
when assuming an entropic prior is equivalent to minimising the function 
\begin{equation}
\Phi_{\rm MEM}({\mathbfss h}) 
= \chi^2({\mathbfss h})-
\alpha S_c({\mathbfss h},{\mathbfss m}_u,{\mathbfss m}_v),
\label{memfunc}
\end{equation}
The minimisation of this $2n_c$-dimensional functions may also
performed using the minimisation algorithm mentioned above, and the
required first derivatives in this case are also given in Appendix A. 

It is important to note that, since we are using the same minimiser to
obtain both the Wiener filter (WF) and MEM reconstructions, and the
evaluation of each function and its derivatives requires similar
amounts of computation, the two methods require approximately the {\em
same} CPU time. Thus, at least in this application, any criticism of
MEM that is based on its greater computational complexity, as compared
to the WF, is no longer valid.
For both the WF and the MEM, the reconstruction of the six $400\times
400$ maps of the input components requires about two minutes on a Sparc
Ultra workstation. 

\subsection{The small fluctuation limit}
\label{slimit}

Despite the formal differences between (\ref{wffunc}) and
(\ref{memfunc}), the WF and MEM approaches are closely related.
Indeed the WF can be viewed as a quadratic approximation to MEM, and
is commonly referred to as such in the literature.
This approximation is most easily verified by considering the small
fluctuation limit, in which the real and imaginary parts of
$\mathbfss{h}$ are small compared to the corresponding models.

Following the discussion at the end of Section \ref{entprior}, 
we begin by setting the real and imaginary parts of all the components of
the models vectors ${\mathbfss m}_u$ and ${\mathbfss m}_v$ 
equal to $m$. Then, expanding the sum in (\ref{entdef}) 
as a power series in $h_j$ and using (\ref{totent}), we find that for 
small $h_j$ the total cross entropy is approximated by
\begin{equation}
S_c({\mathbfss h},{\mathbfss m}_u,{\mathbfss m}_v) 
\approx  -\sum_{j=1}^{n_c} \frac{\Re(h_j)^2+\Im(h_j)^2}{4m}
=  -\frac{{\mathbfss h}^\dagger{\mathbfss h}}{4m}.
\label{entapprox}
\end{equation}
Thus, in the small fluctuation limit, the posterior probability assuming
an entropic prior becomes Gaussian and is given
\begin{equation}
\Pr({\mathbfss h}|{\mathbfss d})
\propto \exp\left[-\chi^2({\mathbfss h})
-\alpha \frac{{\mathbfss h}^\dagger{\mathbfss h}}{4m}\right].
\label{meapprox}
\end{equation}
In fact, this approximation is reasonably accurate provided the
magnitudes of the real and imaginary parts of each element of
$\mathbfss{h}$ are less than about $3m$. Since $m$ is set equal to the
expected rms level of these parameters, we would therefore expect that
for a Gaussian process this approximation should remain  valid. In this
case, the posterior probability (\ref{meapprox}) becomes identical to
that for the WF solution, provided we set $\alpha=4m$. 

We note, however, that for highly non-Gaussian processes,
the magnitudes of the real and imaginary parts of the elements of
$\mathbfss{h}$ can easily exceed $3m$ and in this case the shapes of
the posterior probability for the WF and MEM approaches become
increasingly different.

\subsection{The regularisation constant $\alpha$}
\label{bayesalpha}

A common criticism of MEM has been the arbitrary choice of 
regularisation constant $\alpha$, which is often considered
merely as a Lagrange multiplier. In early applications of MEM,
$\alpha$ was chosen so that the misfit statistic $\chi^2$ equalled
its expectation value, i.e. the number of data points to be fitted.
This choice is usually referred to as historic MEM. 

In the reconstruction of Fourier modes presented here, the situation
is eased somewhat since the choice $\alpha=4m$ is at least guaranteed
to reproduce the results of the Wiener filter when applied to
Gaussian processes. In fact, when applied to the simulations presented
in Section \ref{simobs}, 
this choice of $\alpha$ does indeed bring $\chi^2$ into its
expected statistical range $n_f\pm\sqrt{2n_f}$, where $n_f$ is the
number of (complex) values in the data vector $\mathbfss{d}$ at each
Fourier mode.

Nevertheless, it is possible to determine the
appropriate value for $\alpha$ in a fully Bayesian manner (Skilling
1989; Gull \& Skilling 1990) by
simply treating it as another parameter in our hypothesis space.
It may be shown (see Appendix B) that $\alpha$ must be a solution of
\begin{equation}
-\alpha S_c(\hat{\mathbfss h},{\mathbfss m}_u,{\mathbfss m}_v)
= n_c-\alpha {\rm Tr}({\mathbfss M}^{-1}),
\label{crit}
\end{equation}
where $\hat{\mathbfss h}$ is the hidden vector that maximises the
posterior probability for this value of $\alpha$. The $n_c\times n_c$
matrix ${\mathbfss M}$ is given by
\[
{\mathbfss M} = {\mathbfss G}^{-1/2}{\mathbfss H}_{\rm MEM}{\mathbfss
G}^{-1/2},
\]
where ${\mathbfss H}_{\rm MEM}$ 
is the Hessian matrix of the function $\Phi_{\rm MEM}$
at the point $\hat{\mathbfss{h}}$
and $\mathbfss{G}$ is the metric on image-space at this point.

It should be noted that both the reconstruction $\hat{\mathbfss{h}}$
and the matrix ${\mathbfss M}$ depend on $\alpha$ and so (\ref{crit})
must be solved numerically using an iterative technique such as 
linear interpolation or the
Newton-Raphson method. We take $\alpha=4m$ as our initial estimate in
order to coincide with the Wiener filter in the small fluctuation
limit. For any particular value of $\alpha$, the corresponding
reconstruction $\hat{\mathbfss h}(\alpha)$ is obtained by minimising
$\Phi_{\rm MEM}$ as given in (\ref{memfunc}), and the Hessian of the
posterior probability at this point is then calculated (see Appendix A).
This in turn allows the evaluation of $S_c(\hat{\mathbfss
h},{\mathbfss m}_u,{\mathbfss m}_v)$ and ${\rm Tr}({\mathbfss
M}^{-1})$ respectively. Typically, fewer than ten iterations are
needed in order to converge on a solution $\hat{\alpha}$ 
that satisfies (\ref{crit}).

\subsection{Updating the ICF and models}
\label{update}

In the MEM approach, after the 
Bayesian value $\hat{\alpha}$ for the regularisation constant
has been found, the corresponding posterior probability distribution
is maximised to obtain the reconstruction $\hat{\mathbfss
h}(\hat{\alpha})$, from which the estimate of the signal vector 
$\hat{\mathbfss s}$ may be straightforwardly derived.
Once this has been performed for each Fourier mode, the
reconstruction of the sky emission due to each physical component is
then found by performing an inverse Fourier transform.

We could, of course, end our analysis at this point and use the maps
obtained as our final reconstructions. However, we find that the
results can be further improved by using the current reconstruction to
update the ICF matrix $\mathbfss{L}$ and the models ${\mathbfss m}_u$
and ${\mathbfss m}_v$, and then repeating the entire MEM analysis
discussed above.  At each Fourier mode, the updated models are taken
directly from the current reconstruction and the updated ICF matrix is
obtained by calculating a new signal covariance matrix $\mathbfss{C}$
from the current reconstruction and performing a Cholesky decomposition.
These quantities are then used in the next iteration of the MEM and
the process is repeated until it converges on a final
reconstruction. Usually, fewer than ten such iterations are
required in order to achieve convergence.

We might expect that a similar method may be used in the WF case, by
repeatedly calculating an updated signal covariance matrix from the 
current reconstruction and using it in the subsequent iteration
of the WF analysis. It is well-known, however, that, since the WF
tends to suppress power at higher Fourier modes, the solution
gradually tends to zero as more iterations are performed. One would
thus first have to correct the derived component power spectra in
order to obtain an
unbiased estimator of the real spectrum (since one knows by how much
power has been suppressed, see the discussion in 
5.1) before performing the next iteration. This could somewhat improve
the determination of the dominant processes (but in that case the
exact input spectra is of little impact) but it would not help
with the spectrum determination of the weak processes (since WF
essentially sets their power spectra level at the input level). In
fact WF should rather be thought of as the last `polishing' step of
a component separation, once a first determination of the power
spectra has been achieved by other means (e.g. by singular value
decomposition). Bouchet \& Gispert (1998) have assessed the reachable
accuracy level for the Planck Surveyor in that case. In the following, we shall
restrict our discussion
to the two extreme cases of exact prior knowledge of the
covariance matrix or the prior knowledge of the rms levels only (the power
spectra being all assumed to be white noise). Of course, WF makes much
more sense if the assumed prior is not far from the truth, since it is
designed to take advantage of that information.

\subsection{Estimating errors on the reconstruction}
\label{esterrors}

Once the final reconstruction has been obtained, it is important to be able
to characterise the errors associated with it.  In the case of the
Wiener filter, the reconstructed signal vector $\hat{\mathbfss s}$ 
at each Fourier mode 
may be obtained in a linear manner from the observed data vector using
(\ref{wfrecon}). Thus
the propagation of errors is straightforward and the covariance matrix
of the reconstruction errors at each Fourier mode is given by
(\ref{wferrors}). 

As mentioned in Section \ref{gaussprior}, 
however, this simple propagation of
errors is entirely a result of the assumption of a Gaussian prior, which,
together with the assumption of Gaussian noise, leads to a Gaussian
posterior probability distribution. In terms of the vector of hidden
variables $\mathbfss{h}$ the posterior probability for the WF 
is given by
\[
\Pr({\mathbfss h}|{\mathbfss d})
\propto \exp\left[-\Phi_{\rm WF}({\mathbfss h})\right]
= \exp\left[-({\mathbfss h}-{\hat{\mathbfss h}})^\dagger
{\mathbfss H}_{\rm WF}({\mathbfss h}-{\hat{\mathbfss h}})\right],
\]
where the Hessian matrix ${\mathbfss H}_{\rm WF}$ is given by
${\mathbfss H}_{\rm WF} = \nabla_{\mathbfss h}\nabla_{{\mathbfss
h}^*}\Phi_{\rm WF}$ evaluated at the peak $\hat{\mathbfss h}$ of the
distribution, and the function $\Phi_{\rm WF}$ is given by
(\ref{wffunc}).  Thus, the covariance matrix of the errors on the
reconstructed hidden vector is then given {\em exactly} by the inverse
of this matrix, i.e
\[
\langle 
({\mathbfss h}-\hat{\mathbfss h})({\mathbfss h}-\hat{\mathbfss h})^\dagger
\rangle 
={\mathbfss H}_{\rm WF}^{-1}.
\]
From (\ref{icfdef}), the error covariance matrix for the reconstructed
signal vector is then given by
\begin{equation}
\langle 
({\mathbfss s}-\hat{\mathbfss s})({\mathbfss s}-\hat{\mathbfss s})^\dagger
\rangle
=
\langle 
{\mathbfss L}
({\mathbfss h}-\hat{\mathbfss h})({\mathbfss h}-\hat{\mathbfss h})^\dagger
{\mathbfss L}^{\rm T}
\rangle
={\mathbfss L}{\mathbfss H}_{\rm WF}^{-1}{\mathbfss L}^{\rm T}.
\label{errgen}
\end{equation}
Using the expression for the Hessian matrix given in (\ref{wfhess}),
and remembering that $\mathbfss{s}=\mathbfss{Lh}$ and ${\mathbfss C} =
{\mathbfss L}{\mathbfss L}^{\rm T}$, the expression (\ref{errgen}) is
easily shown to be identical to the result (\ref{wferrors}).

For the entropic prior, the posterior probability
distribution is not strictly Gaussian in shape. Nevertheless, we may
still approximate the shape of this distribution 
by a Gaussian at its maximum and, 
recalling the discussion of subsection \ref{slimit}, we might
expect this approximation to be reasonably accurate, particularly in 
the reconstruction of Gaussian processes. Thus, near the point
$\hat{\mathbfss h}$, we make the approximation
\[
\Pr({\mathbfss h}|{\mathbfss d})
\propto \exp\left[-\Phi_{\rm MEM}({\mathbfss h})\right]
\approx \exp\left[-({\mathbfss h}-{\hat{\mathbfss h}})^\dagger
{\mathbfss H}_{\rm MEM}({\mathbfss h}-{\hat{\mathbfss h}})\right],
\]
where ${\mathbfss H}_{\rm MEM} = \nabla_{\mathbfss h}\nabla_{{\mathbfss
h}^*}\Phi_{\rm MEM}$ evaluated at $\hat{\mathbfss h}$, and
$\Phi_{\rm MEM}$ is given by (\ref{memfunc}).
The covariance matrix of the errors on the reconstructed hidden vector
is then given {\em approximately} by the inverse of this matrix, and
so
\[
\langle 
({\mathbfss s}-\hat{\mathbfss s})({\mathbfss s}-\hat{\mathbfss s})^\dagger
\rangle
\approx 
{\mathbfss L}{\mathbfss H}_{\rm MEM}^{-1}{\mathbfss L}^{\rm T}.
\]

In both the WF and MEM cases, the reconstructed maps of the
sky emission due to each physical component is obtained by inverse
Fourier transformation of the signal vectors at each Fourier
mode. Since this operation is linear, the errors on these maps may
therefore be deduced straightforwardly from the above error covariance
matrices.

\section{Application to simulated observations}
\label{results}

We now apply the MEM and WF analyses outlined above to the simulated
Planck Surveyor data discussed in Section \ref{simobs}. Clearly, both
techniques rely to some extent on our prior knowledge of the input
components we are trying to reconstruct. Information concerning
the spectral behaviour of each component is contained in the frequency
response matrix $\mathbfss{F}$ defined in (\ref{freqresp}), whereas the 
assumed covariance structure of the components is
contained in the signal covariance matrix $\mathbfss{C}$ given in
(\ref{covdef}). Since we are in fact performing the reconstruction in the
Fourier domain, the latter matrix contains the power spectrum of
each component as its diagonal entries and the cross power spectra
between components as its off-diagonal entries. Strictly speaking,
since we reconstruct the vector of hidden variables $\mathbfss{h}$,
rather than the signal vector $\mathbfss{s}$, this power spectrum 
information actually resides in the ICF matrix $\mathbfss{L}$.

For the reconstructions presented in this section, we assume that
the spectral behaviour of the components is accurately known.
This is certainly true for the CMBR emission and the kinetic and
thermal SZ effects, but it is perhaps optimistic to assume 
that this would be the case for the three Galactic components.
In reality the spectral indices of the free-free and synchrotron emission 
are uncertain to within about 20 per cent and the dust temperature
and emissivity will also not be known in advance. We have investigated
the effect of varying these parameters in the reconstruction algorithms
and have found both the MEM and WF separations to be quite robust.
This is discussed further in Section \ref{conc} (see also Gispert \&
Bouchet 1997).

Our prior knowledge of the covariance structure or power spectra of
the various emission components is certainly poorer than our knowledge
of their spectral behaviour. Nevertheless, we are not entirely
ignorant of the shapes of these power spectra and we would
obviously wish to include any such information in our analysis. In
order to investigate how the quality of the reconstructions depends on
our knowledge of the power spectra, we have chosen to model two
extreme cases. First, we assume knowledge of the azimuthally-averaged
power spectra of all six input components, as shown in
Fig.~\ref{fig2}, together with the azimuthally-averaged cross power
spectra between components; these contain
cross-correlation information in Fourier space, so that
the ICF matrix $\mathbfss{L}$ is fully specified. In the
second case, however, we take the opposite view and assume that almost
no power spectrum information is available. This corresponds to
assuming a flat (white-noise) power spectrum for each component out to
the highest measured Fourier mode. The levels of the flat power
spectra are chosen so that the total power in each component is
approximately that observed in the input maps in Fig.~\ref{fig1}.

It is likely, in practice, that the quality of prior information
concerning the component power spectra will lie somewhere between
these two extreme cases. Therefore, by presenting the results for each
case, we aim to provide some idea of the best- and worst-case limits
on the quality of component separation that can be achieved for the
Planck Surveyor mission. In addition, we hope to illustrate the
different behaviour of the MEM and WF techniques in the two extreme
r\'egimes. The reconstructions with full ICF information are intended
to display that the main advantage of the MEM technique in this case
is its superiority in reconstructing weak non-Gaussian processes. In
the absence of ICF information, we wish to illustrate that the
iterative formulation of MEM presented above allows the component
separation to be nearly as efficient without prior knowledge as it is
when the ICF matrix is fully specified.  On the other hand, we
anticipate much larger differences between the two r\'egimes for the
WF reconstructions, since this method amounts to designing optimal
linear filters by using prior knowledge of the component power
spectra.  Nevertheless, although WF makes more sense as a method when
the assumed prior information is close to truth, it is of interest to
investigate the robustness the reconstructions of the various
components in the absence of such information.  Indeed, if the CMB is
well reconstructed with essentially no prior information given to the
WF, then its estimate is truly very robust.

\subsection{Reconstructions with full ICF information}
\label{recicf}

We first consider the case in which power spectrum and
cross-correlation information are assumed, so that the ICF matrix
$\mathbfss{L}$ is fully specified. 
In this case, the Bayesian value
of the regularising parameter $\alpha$ that satisfies (\ref{crit}) 
is found to be $\alpha = 0.8$.

\subsubsection{The reconstructed maps}

\begin{figure*}
\centerline{\epsfig{
file=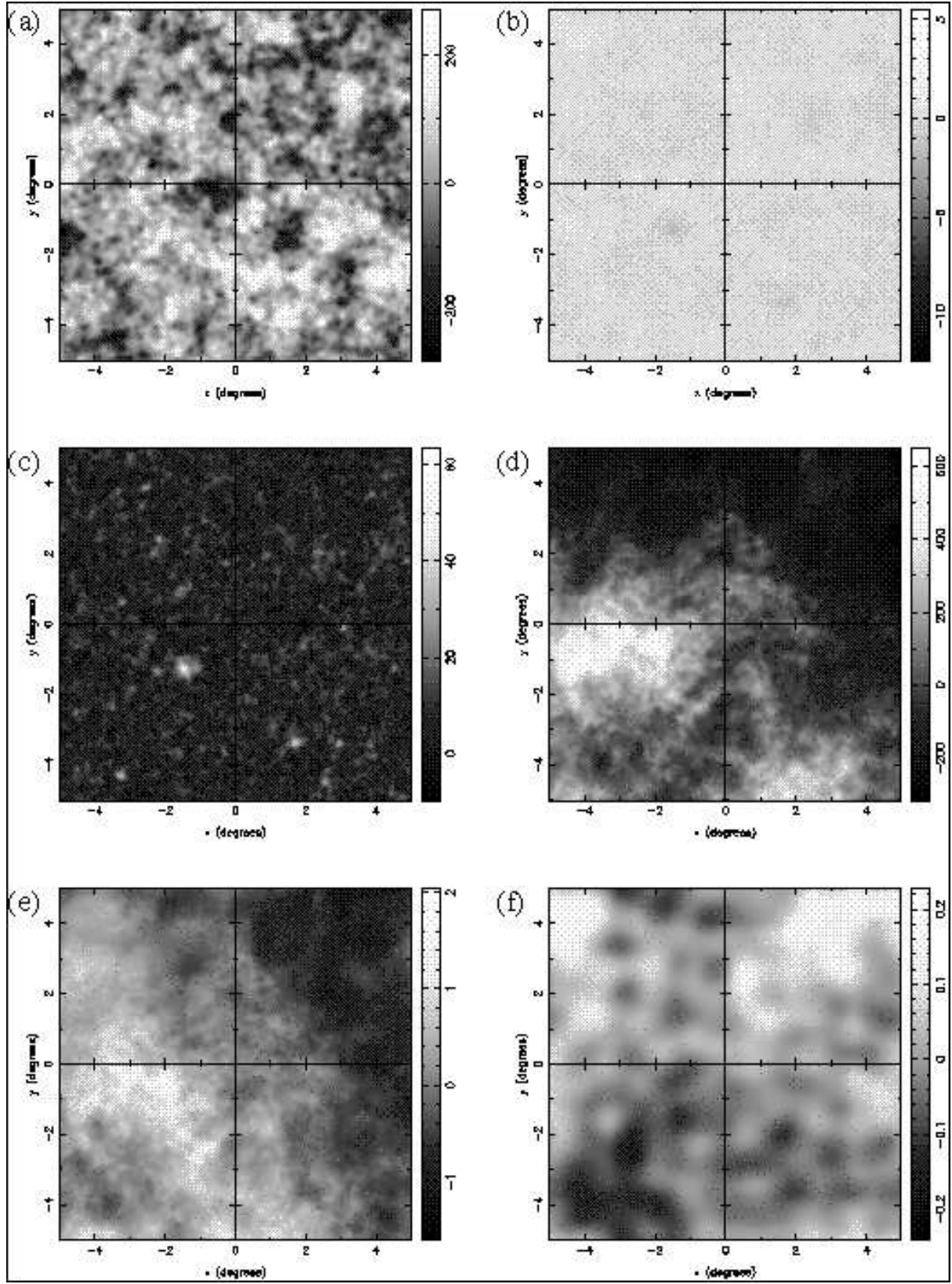,width=16cm}}
\caption{MEM reconstruction of the $10\times 10$-degree maps of the
six input components shown in Fig.~\ref{fig1}, using full ICF
information (see text). The components are: (a) primary CMBR
fluctuations; (b) kinetic SZ effect; (c) thermal SZ effect; (d)
Galactic dust; (e) Galactic free-free; (f) Galactic synchrotron
emission.  Each component is plotted at 300 GHz and has been convolved
with a Gaussian beam of FWHM equal to 4.5 arcmin. The map units are
equivalent thermodynamic temperature in $\mu$K.}
\label{fig5}
\end{figure*}

\begin{figure*}
\centerline{\epsfig{
file=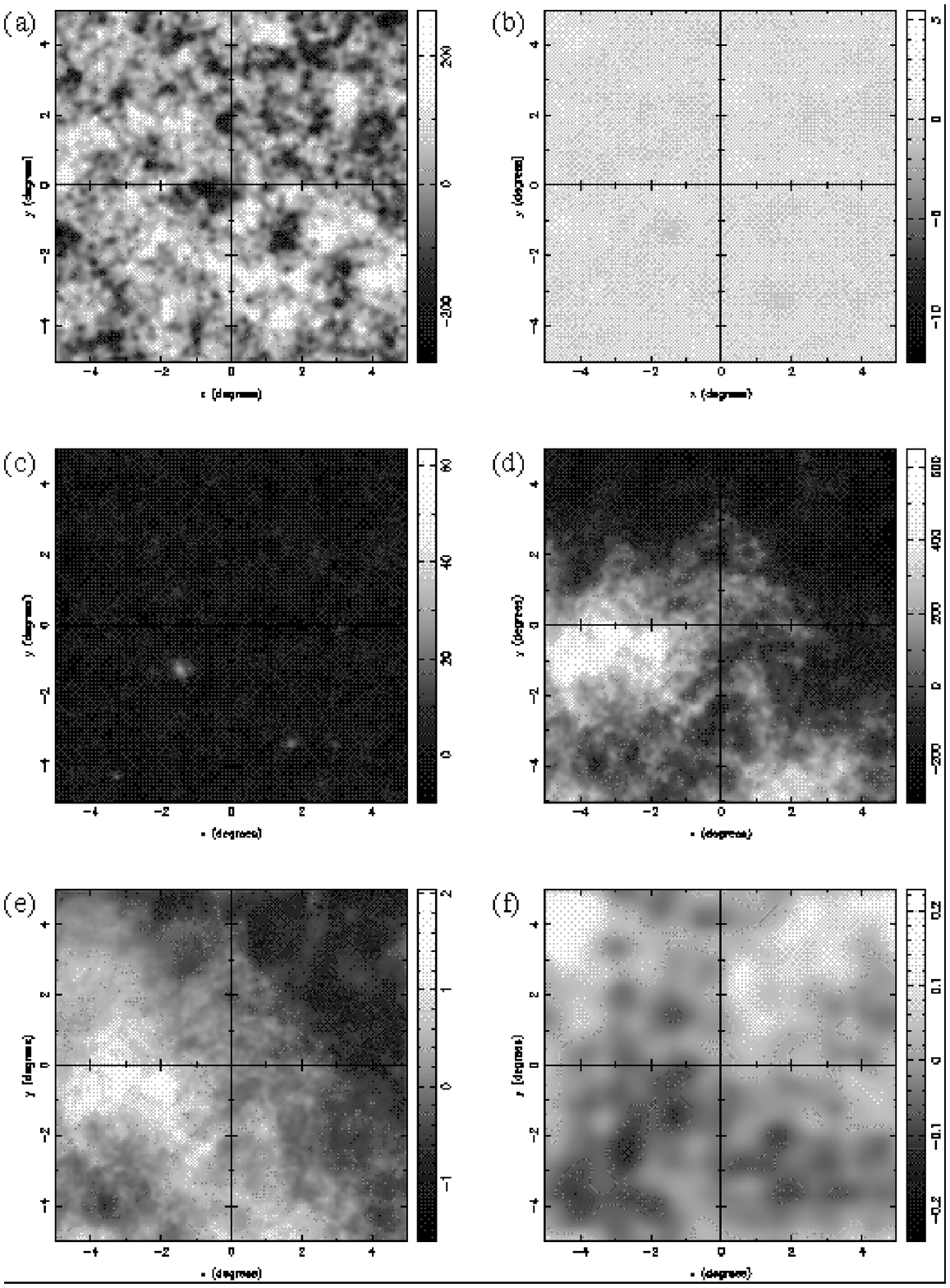,width=16cm}}
\caption{Wiener filter reconstruction of the $10\times 10$-degree maps of
the six input components shown in Fig.~\ref{fig5}, using full ICF
information (see text).}
\label{fig6}
\end{figure*}

The corresponding MEM and WF reconstructions of the six
input components shown are shown in Figs~\ref{fig5} and \ref{fig6}
respectively, convolved with a 4.5 arcmin FWHM Gaussian beam.
The grey scales in these figures are chosen to coincide
with those in Fig.~\ref{fig1} in order to enable a more
straightforward comparison with the input maps.

We see that the main input components are
faithfully reconstructed. Perhaps most importantly, the CMBR has been
reproduced extremely accurately, and at least by eye both the MEM and
WF reconstructions are virtually indistinguishable from the true input
map. As we might expect the dust emission is also accurately
recovered, since it dominates the high frequency channels.  The
free-free emission, which is highly correlated with the dust, has also
been reconstructed well, with both the MEM and WF
reconstructions containing most of the main features present in the
true input map. 
The recovery of the synchrotron emission is also reasonable, although
the MEM algorithm is more successful in recovering the brightest
regions.

The MEM and WF reconstructions of the kinetic and thermal SZ effects
are worth some comment. Both techniques have produced reasonable
reconstructions of the thermal SZ effect for those clusters in which
the effect is very strong. However, it is clear that the MEM has
successfully reconstructed the SZ effect in a greater number of
clusters. Moreover, the magnitudes of the SZ effects in the MEM
reconstruction are closer to the true values than those obtained with
the WF. Thus, as anticipated, the assumption of Gaussian random fields
that is central to the WF approach leads to poorer reconstructions of
highly non-Gaussian fields as compared with MEM. A more detailed
discussion of the recovery of thermal SZ profiles is given in Section
\ref{szprof}. For the kinetic SZ effect, however, neither method has
reconstructed any features in the input map with their
true magnitudes. In fact,
both methods have reconstructed fields with very low-level
fluctuations that coincide with the brightest features in the thermal
SZ map. 
The inability of either method to make very accurate reconstructions
of the kinetic SZ effect is not
surprising since, as mentioned in Section \ref{simobs}, 
this emission due to this component is at least two orders of
magnitude below the dominant emission component or the noise
at all of the Planck Surveyor observing frequencies. Moreover,
this component has the same frequency dependence as the primordial
CMBR fluctuations, and so we cannot distinguish them by their spectral
behaviour. Nevertheless, it is still possible to distinguish between
the CMBR and kinetic SZ emission on the basis of their different power
spectra, and we do indeed obtain 
marginal detections of the kinetic SZ effect in some clusters;
this is also discussed in Section \ref{szprof}.

While a visual inspection of the reconstructed maps is a useful method
of assessing how well the algorithms are performing, a more
quantitative analysis of the reconstruction errors is required if we
are to make any meaningful comparison between the MEM and WF
approaches. The most straightforward means of comparison is to
calculate the rms of the residuals for each set of reconstructions.
For any particular physical component, this is given by
\[
e_{\rm rms} = \left[\frac{1}{N}\sum_{i=1}^N 
\left(T_{\rm rec}(\bmath{x}_i) - T_{\rm true}(\bmath{x}_i)\right)^2
\right]^{1/2},
\]
where $T_{\rm rec}(\bmath{x}_i)$ and $T_{\rm true}(\bmath{x}_i)$
are respectively the reconstructed and true temperatures in the $i$th
pixel; $N$ is the total number of pixels in the map. The values
of $e_{\rm rms}$ for each physical component in the MEM and WF 
reconstructions are shown in Table~\ref{erms1}. 
Since, for comparison purposes, both the
input and reconstructed maps have been convolved with a 4.5 arcmin
FHWM Gaussian, the $e_{\rm rms}$ values quoted should interpreted as
the rms residual per beam of this size.
\begin{table}
\begin{center}
\caption{The rms residuals per 4.5 arcmin FHWM Gaussian beam
(in $\mu$K) for the MEM and WF
reconstructions shown in Figs~\ref{fig5} and \ref{fig6}, which
assume full ICF information.}
\begin{tabular}{lcc} \hline
Component & $e^{\rm MEM}_{\rm rms}$ & $e^{\rm WF}_{\rm rms}$ \\ \hline
CMBR             & 5.90 & 6.00 \\
Kinetic SZ       & 0.85 & 0.85 \\
Thermal SZ       & 3.90 & 4.10 \\
Dust             & 1.60 & 1.90 \\
Free-Free        & 0.30 & 0.37 \\
Synchrotron      & 0.05 & 0.06 \\ \hline
\end{tabular}
\end{center}
\label{erms1}
\end{table}
We see from the table that, in terms of the rms of the reconstruction
errors, the two methods are nearly equivalent. In particular, we note
that the CMBR has been reconstructed to an accuracy of about 6$\mu$K,
which is the desired value quoted for the Planck Surveyor mission
(Bersanelli et al. 1996). We note, however, that the rms error for MEM
reconstruction is slightly smaller than for the WF.  The rms errors
for the other components are also similar for the MEM and WF
reconstructions, but are always lower for the MEM algorithm.
This is particularly true for the reconstructions of the thermal
SZ effect, dust and free-free emission and is 
due in part to the non-Gaussian nature of these components.

\begin{figure*}
\centerline{\epsfig{
file=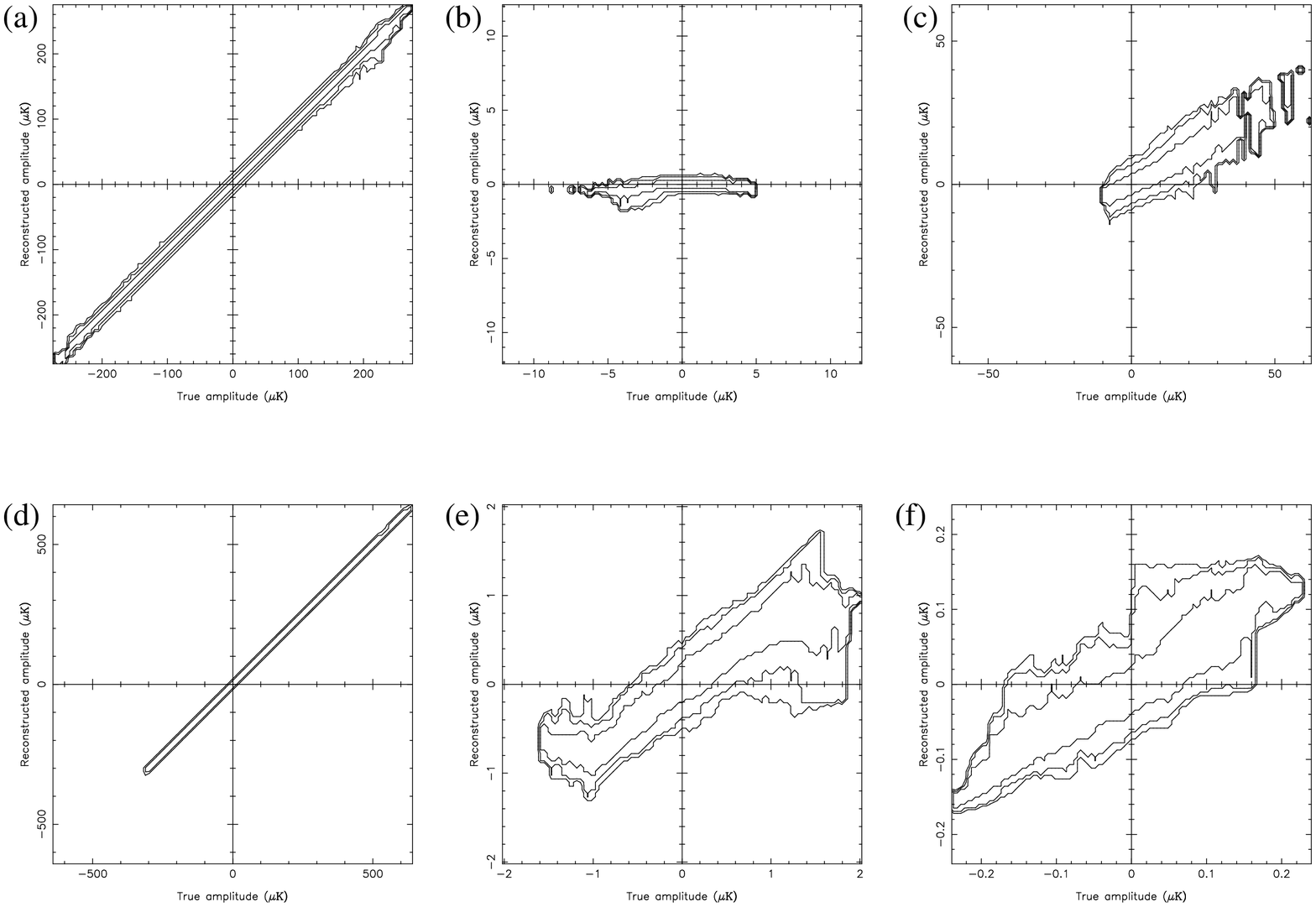,width=15.0cm}}
\caption{Comparison of the input maps with the maps reconstructed 
using the MEM algorithm with full ICF information.
The horizontal axes show the input map amplitude within a pixel 
and the vertical axes show the reconstructed amplitude. 
The contours contain 50 and 99 per cent of the pixels respectively.}
\label{fig7}
\end{figure*}

\begin{figure*}
\centerline{\epsfig{
file=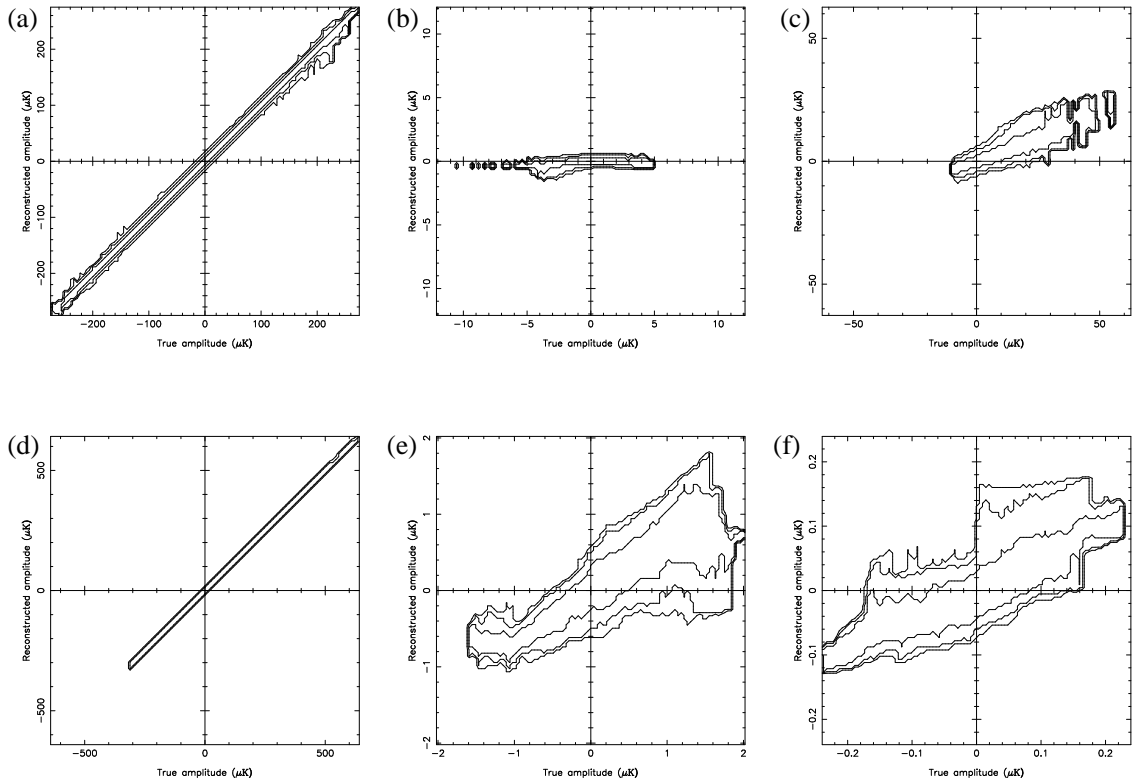,width=15.0cm}}
\caption{As for Fig.~\ref{fig7} but for the Wiener filter
reconstruction with full ICF information.}
\label{fig8}
\end{figure*}

Simply quoting the rms of the residuals is, however, a
rather crude method of quantifying the accuracy of the
reconstructions. A more useful approach is to characterise the
reconstruction errors on a given component by plotting the amplitudes
of the temperature fluctuations for each pixel of the reconstructed
map against those in the true map. Usually such plots consist of a
collection of points, one for each pixel in the true/reconstructed
map. We shall, however, adopt a slightly different approach 
For each component, the temperature range of the true map is divided
into 100 bins. Three contours are then plotted which
correspond to the 68, 95 and 99 per cent
points of the distribution of corresponding reconstructed
temperatures in each bin. If the reconstruction is
particularly good, then only the 95 and 99 per cent contours are
plotted. Clearly, a perfect reconstruction would be represented by a
single diagonal box of width equal to the bin size used.
Figs~\ref{fig7} and \ref{fig8} show the comparison plots for the WF
and MEM reconstructions respectively and Table~\ref{slope1} gives
the gradient of the best-fit straight line through the origin
for each component.


Panel (a) in each figure shows the confidence limits for the
reconstruction of the CMBR, and it is clear that both reconstructions
are very accurate.  In each case, for those points in the true CMBR
map with temperatures lying in the range $-200$ $\mu$K to 200 $\mu$K,
the 68 per cent limits of the reconstructed temperatures lie
approximately 5$\mu$K on either side of the true value. This agrees
with the values for $e_{\rm rms}$ for each reconstruction, given in
Table~\ref{erms1}. For points in the true map having very large
positive or negative values, both the MEM and WF reconstructions
become slightly less accurate, but the errors are still in the range
5--10 $\mu$K.  Comparing the performance of the MEM and WF approaches,
there is some evidence in that the MEM reconstruction is slightly more
accurate for points having large positive temperatures, and it is
these points that make the largest contribution to difference in the
values of $e_{\rm rms}$ given in Table.~\ref{erms1}.

Panels (b) and (c) in Figs~\ref{fig7} and \ref{fig8} show the
confidence limits for the reconstruction of the kinetic and thermal SZ
respectively. As we would anticipate from the maps of the kinetic SZ
reconstructions in Figs~\ref{fig5}(b) and \ref{fig6}(b), for both the
MEM and WF techniques, the distribution of the reconstructed
temperatures centres around zero for all values of the temperature in
the input map.  For the thermal SZ, however, we see that the
reconstructions are considerable better. Nevertheless, for both
reconstructions, the best-fit straight line through the origin has a
slope that is significantly smaller than unity, indicating that
magnitudes of the thermal SZ effects are generally underestimated. It
is clear from the plots that this effect is more pronounced in the WF
reconstruction, since the corresponding best-fit line has a markedly
lower slope than for the MEM reconstruction. About the corresponding
best-fit line the range in the values of the reconstructed
temperatures is slightly smaller for the WF reconstruction than for
MEM. However, the bias in the best-fit line and the relatively low
dispersion in the WF case are both due to its signal-to-noise
weighting to reach minimum variance estimates. This tends
to reduce the reconstructed values
for weak processes like the thermal SZ. Accounting for
the bias (see Section \ref{conc}
for further discussion) would boost the recovered
mode values and decrease the bias while keeping constant the
signal-to-noise, i.e. it would result in more noise in the recovered
maps without changing the significance level of the detections. We
have not done performed this correction and have instead
kept the standard minimum-variance WF procedure. In
this case, the standard deviation of the reconstructed temperatures
about the {\em true} temperature is lower for MEM, as indicated by the
relatives values of $e_{\rm rms}$ for this component given in
Table~\ref{erms1}.  The tendency for both methods to underestimate
the magnitude of the thermal SZ effects is due to the fact that the
emission in this component is dominated by dust emission and pixel
noise at the observing frequencies with the highest angular
resolutions. Thus information concerning the higher Fourier modes of
the thermal SZ map is not present in the data and so very sharp
features are unavoidably smoothed in the reconstructions.

The confidence limits for the reconstructions of the Galactic
components are shown in panels (d), (e) and (f) of Figs~\ref{fig7} and
\ref{fig8}; these correspond to dust, free-free and synchrotron
emission respectively. The confidence contours for the MEM and WF
reconstructions of the dust component are indistinguishable and
clearly show that the dust is the most accurately reconstructed
component. The 99 per cent limits of the reconstructed temperature
distributions are approximately constant for all values of the true
input temperature and correspond to 3$\sigma$ error in the
reconstruction of about 5$\mu$K. From panels (e) and (f) it is clear
that the reconstructions of the free-free and synchrotron emission are
considerable less accurate. By comparing these plots for the MEM and
WF reconstructions, we again notice (for the
same reasons as noted above) that the best-fit straight line
through the origin has a slope that is closer to unity for MEM than
for the WF and the standard deviation of the reconstructed
temperatures about these lines is also smaller for MEM, as
indicated by the
smaller corresponding values of $e_{\rm rms}$ in each case .  The
relative large spread of reconstructed temperatures for the free-free
and synchrotron components is due partially to the fact that the
reconstructions have low effective resolution, since the Planck
Surveyor has relatively large beam sizes at the lower observing
frequencies where the free-free and synchrotron emission is
highest. If the input maps are instead convolved to a lower
resolution, such as 20 arcmin, which is more typical of the beam sizes
at the lower observing frequencies, then the spread in the
reconstruction values is considerably reduced. In fact for WF alone, one
should rather convolve the input map with the effective beam of the
reconstruction as determined from the WF method itself (see Bouchet et
al. 1997 for examples of such beams), but this would prevent a
straight comparison with MEM. 

\begin{table}
\begin{center}
\caption{The gradients of the best-fit straight line through the
origin for the comparison plots shown in Figs ~\ref{fig7} and
\ref{fig8} for the MEM and WF reconstructions, which assume full 
ICF information.}
\begin{tabular}{lcc} \hline
Component        &   MEM gradient    &    WF gradient     \\ \hline
CMBR             & $1.00$  &  $1.00$  \\
Kinetic SZ       & $0.06$  &  $0.05$  \\
Thermal SZ       & $0.55$  &  $0.27$  \\
Dust             & $1.00$  &  $1.00$  \\
Free-Free        & $0.48$  &  $0.37$  \\
Synchrotron      & $0.62$  &  $0.44$  \\ \hline
\end{tabular}
\end{center}
\label{slope1}
\end{table}

\subsubsection{The reconstructed power spectra}
\label{error}

\begin{figure*}
\centerline{\epsfig{
file=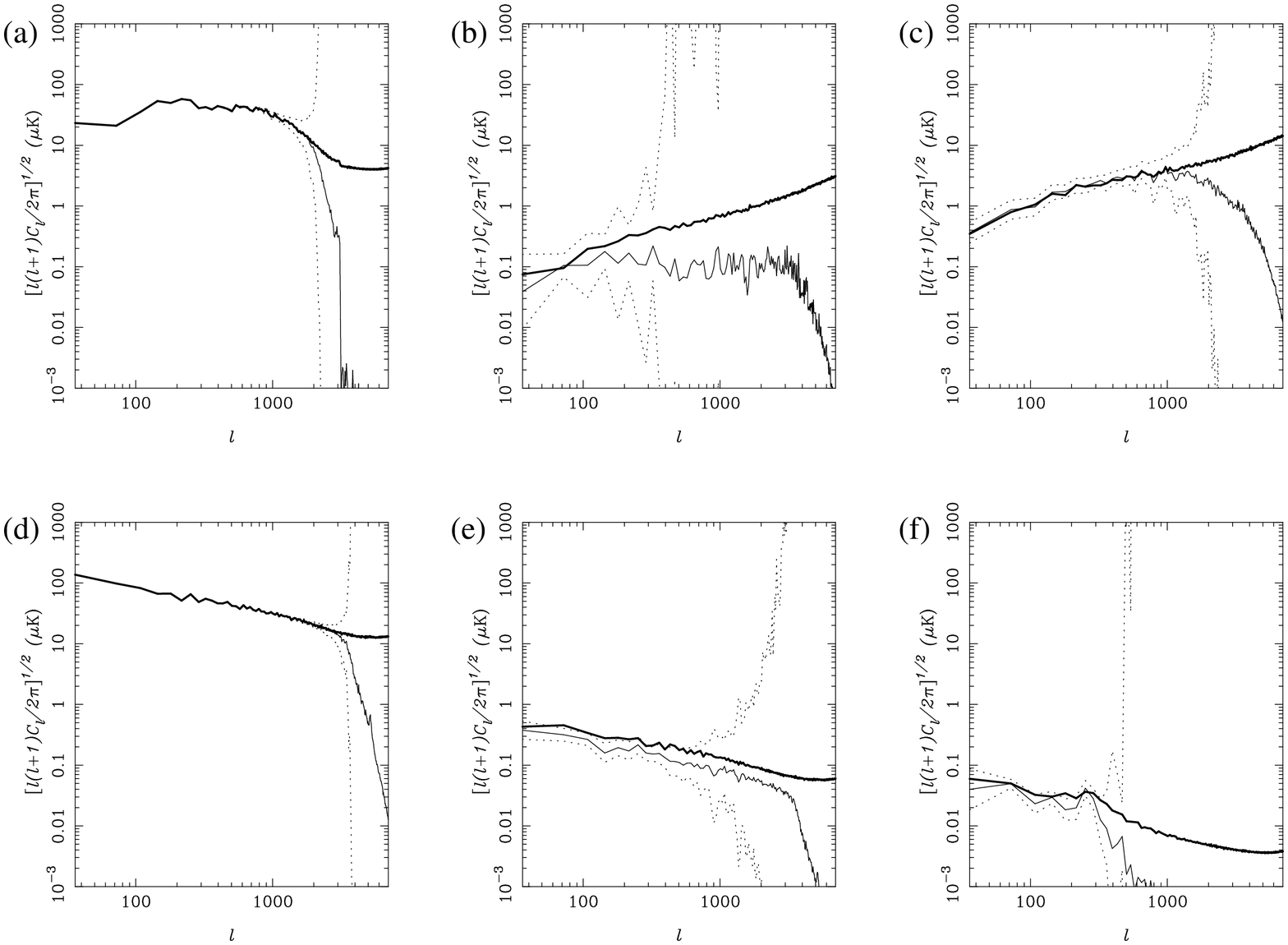,width=14.5cm}}
\caption{The power spectra of the input maps (bold line) compared to
to the power spectra of the maps reconstructed using MEM with full ICF
information (faint line). The dotted lines show one sigma confidence
limits on the reconstructed power spectra.}
\label{fig9}
\end{figure*}
\begin{figure*}
\centerline{\epsfig{
file=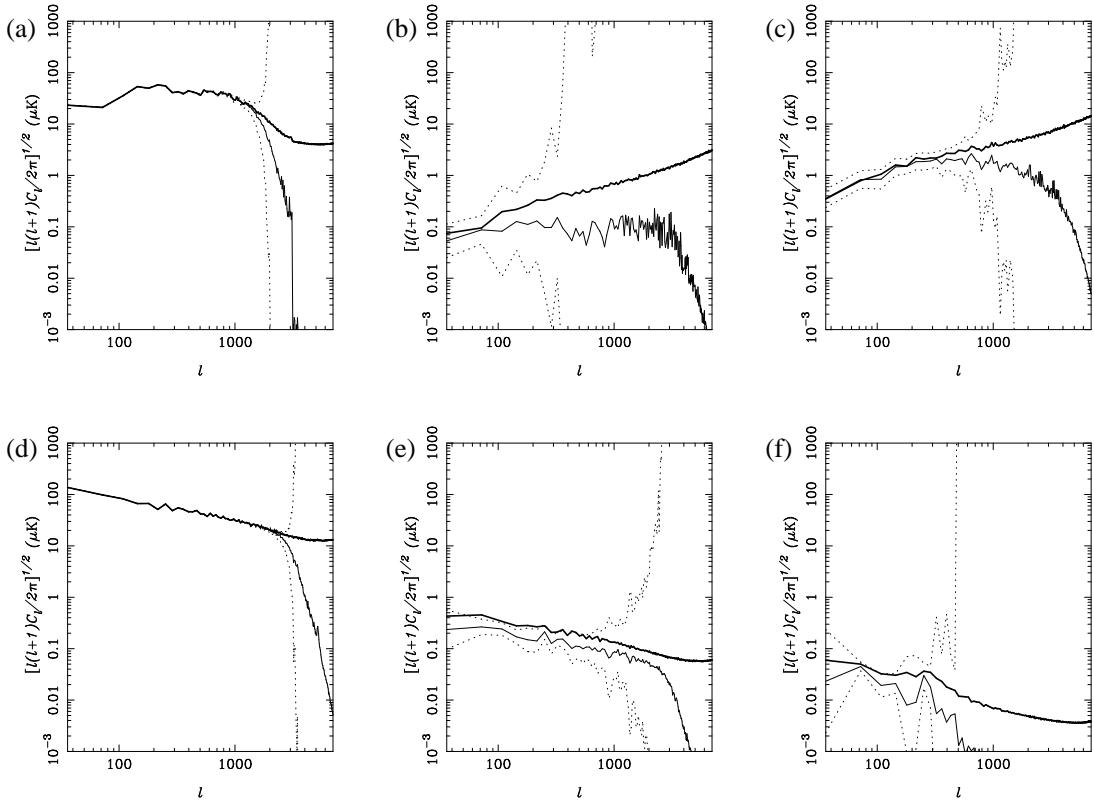,width=14.5cm}}
\caption{As for Fig~\ref{fig9}, but for the WF reconstruction with
full ICF information.}
\label{fig10}
\end{figure*}

Since both the MEM and WF reconstructions are performed in the Fourier
domain, it is particularly straightforward to compute the
reconstructed power spectra of the physical components. Both
techniques reconstruct the signal vector $\hat{\mathbfss s}(\bmath{k})$ 
at all measured Fourier modes. These Fourier modes lie on a square
$400\times 400$ grid with a grid spacing $\Delta k = 36$ wavenumbers. 
At a given value of $k$, the estimator $\hat{C}_p(k)$ of the
azimuthally averaged power spectrum 
for the $p$th physical component is obtained simply by calculating the 
average value of $|\hat{s}_p(\bmath{k})|^2$ over those modes for which
$|\bmath{k}|=k$, i.e.
\begin{equation}
\hat{C}_p(k) = \frac{1}{N(k)}
\sum_{|\bmath{k}_i|=k} \hat{s}_p(\bmath{k}_i)\hat{s}^*_p(\bmath{k}_i),
\label{psest}
\end{equation}
where $N(k)$ is the number of measured Fourier modes satisfying
$|\bmath{k}_i|=k$. We note that for $k \ga 36$ it is a reasonable
approximation to identify the flat two-dimensional wavenumber $k$ with
the spherical harmonic multipole index $\ell$. 
The errors on the reconstructed power spectrum are also easily
estimated from the errors on the reconstructed signal vectors at each
Fourier mode. It is straightforward to show that
\begin{eqnarray*}
{\rm Var}[\hat{C}_p(k)] 
& \approx & 2 \sum_{|\bmath{k}_i|=k} 
\pd{\hat{C}_p(k)}{\hat{s}_p(\bmath{k}_i)}
\pd{\hat{C}_p(k)}{\hat{s}^*_p(\bmath{k}_i)}
{\rm Var}[\hat{s}_p(\bmath{k}_i)] \\
& \approx & \frac{2}{N^2(k)}\sum_{|\bmath{k}_i|=k} 
\hat{s}_p(\bmath{k}_i) \hat{s}^*_p(\bmath{k}_i)
{\rm Var}[\hat{s}_p(\bmath{k}_i)].
\end{eqnarray*}

For a WF reconstruction, it is well known that $\hat{C}_p(k)$ is a
biased estimator of the underlying power spectrum (see e.g. Bouchet et
al. 1997).  Nevertheless, this is not necessarily the case for the MEM
reconstruction and, for comparison purposes, it is instructive to use
the same power spectrum estimator for both the MEM and WF
reconstructions. Moreover, in this section we are interested simply in
the power spectra of the reconstructed maps, rather than in developing
optimal methods to recover the input power spectrum from a given
reconstruction.  In Section \ref{conc}, we discuss in more detail the
biased nature of this simple power spectrum estimate, and consider
several variants of the standard Wiener filter that may be used to
circumvent this problem. At this point, however, it is sufficient to
note that where the underlying power spectrum of the $p$th process is
poorly determined by the observations, the estimator
$\hat{C}_p(k)$ can be shown to underestimate the true power spectrum.

\begin{figure*}
\centerline{\epsfig{
file=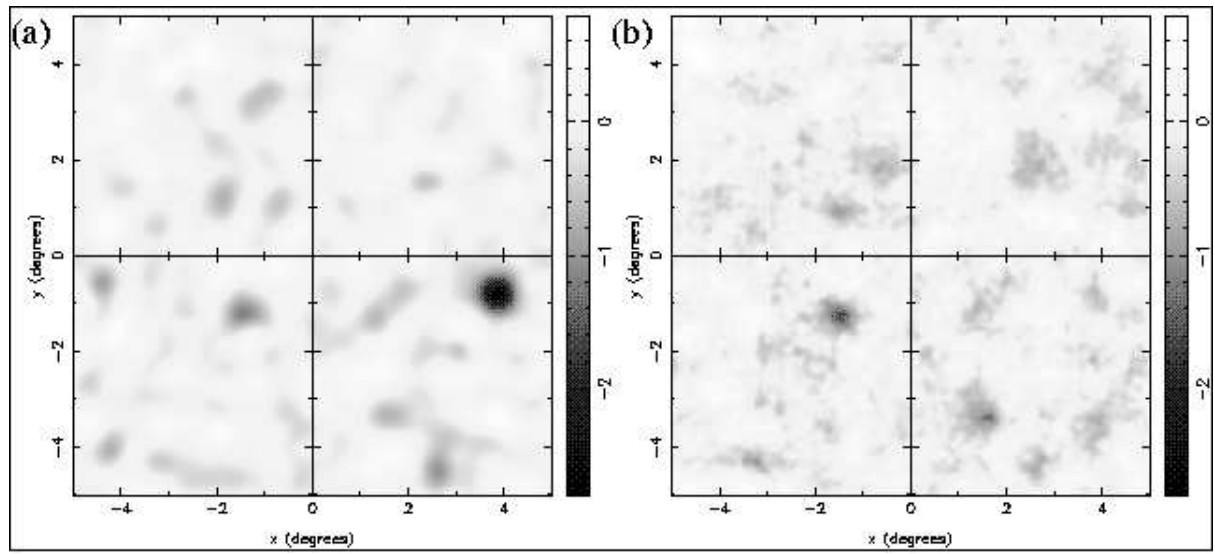,width=16cm}}
\caption{(a) The input kinetic SZ map convolved to the lowest Planck
Surveyor angular resolution of 33 arcmin. (b) The MEM reconstruction
of the kinetic SZ effect. The map units are equivalent thermodynamic
temperature in $\mu$K at 300 GHz.}
\label{fig11}
\end{figure*}

Figs.~\ref{fig9} and \ref{fig10} show the power spectra
of the MEM and WF reconstructions respectively, 
together with the 68 per cent error
bars. In each panel the faint line is the power spectrum of the
reconstructed map and the bold line is the power spectrum of the
relevant input map as shown in Fig.~\ref{fig2} (for the power spectrum
comparison the maps are {\em not} convolved by a 4.5 arcmin FHWM
Gaussian beam). We see that the 68 per
cent confidence intervals always contain the true power spectrum,
which indicates that our estimate of the errors on the reconstructed
power spectrum are quite robust.

The power spectrum of the reconstructed CMBR maps are shown in panel
(a) of each figure, and we see that both techniques have faithfully
reproduced the true power spectrum for $\ell \la 1500$, at which point
the WF reconstructed map begins visibly to underestimate the true
spectrum. The MEM reconstruction, however, remains indistinguishable from the
true power spectrum up to $\ell \approx 2000$, where it too begins to
underestimate the true spectrum. 

The power spectra of MEM and WF reconstructions of the kinetic SZ map
are shown in panel (b) of each figure and are predictably poor, with
both reconstructed power spectra underestimating the true one over
almost the entire range of measured multipoles. For the thermal SZ
component shown in panel (c), both methods produce maps with power
spectra that lie close to the true spectrum for at lower
multipoles. However, we again find that the MEM reconstruction remains
faithful out to larger multipoles ($\ell \approx 1000$) as compared to
the WF reconstruction ($\ell \approx 300$).

Panels (d), (e) and (f) in Fig.~\ref{fig9} and \ref{fig10} show the
power spectra of the MEM and WF reconstructions of the Galactic dust,
free-free and synchrotron. As expected, for the dust component
both methods produce reconstructions with power spectra that are very
close to the power spectrum of the true map over a large range of
multipoles. The power spectrum of the MEM reconstruction is
indistinguishable from that of the true map up to $\ell \approx 3000$,
whereas the WF reconstruction becomes inaccurate at $\ell \approx
2000$. For the free-free component, both MEM and WF produce
reconstructions with power spectra that slightly underestimate the
true spectrum over the entire range of measured multipoles.  Finally,
the power spectra of the synchrotron reconstructions show the MEM
technique reproduces the true power spectrum to moderate accuracy for
$\ell \la 400$, whereas the WF reconstruction underestimates the true
power spectrum for all multipoles.

\subsubsection{The reconstructed kinetic and thermal SZ effects}
\label{szprof}

As discussed in Section \ref{gaussprior}, a central assumption of the
Wiener filter method is that the fields to be reconstructed are well
described by Gaussian statistics. This is clearly not a valid
assumption for either the kinetic or thermal SZ effects for which the
emission consists of sharp peaks. Thus we would expect that it is in
the reconstruction of this component especially that the difference
between the MEM and WF approaches would be most apparent.

Unfortunately, the small magnitude of the kinetic SZ, together with a
frequency spectrum identical to that of the primary CMBR fluctuations,
means that neither of the methods is capable of reconstructing this
component very accurately.
Nevertheless, the both methods do make marginal detections of the
kinetic SZ effect in some clusters. Fig.~\ref{fig11} shows the MEM
reconstruction of the kinetic SZ map compared to the true map
convolved to the lowest Planck Surveyor resolution of 33 arcmin; it is
in this lowest frequency channel that the relative contribution of the
kinetic SZ effect to the total emission is highest. From the figure,
we see that the MEM algorithm has recovered the kinetic SZ effect at
this lowest resolution, but only in a few clusters.  By comparing
these maps with the MEM thermal SZ reconstruction in
Figs~\ref{fig5}(c), we see that these clusters are those with the
largest thermal SZ effects. Conversely, the largest kinetic SZ effect
in the true map is not recovered with any accuracy, since by chance it
corresponds to a cluster with a small thermal SZ effect.

For the thermal SZ, we see from Figs~\ref{fig5}(c) and
\ref{fig6}(c) that both the MEM and WF algorithms reproduce the main
features present in the input map, but that MEM reconstructs the
thermal SZ effect in many more clusters than the WF and that the
magnitude of the reconstructed effects using MEM are closer to those
in the input map. This observation is confirmed by investigating the
errors on the reconstructed maps and by comparing the power spectra of
the input map and the reconstructions.

It is hoped that Planck Surveyor observations of the thermal SZ
effect, together with follow-up X-ray observations of the relevant
galaxy clusters, will provide a large catalogue of $H_0$
determinations to supplement the value of $H_0$ obtained from the
accurate measurement of the primordial CMBR power spectrum. In order
for this to be possible, however, the density profile of the clusters
must be known.  Furthermore, an accurate determination of the density
profile of a cluster enables the construction of optimal filters,
tuned to the individual cluster characteristics, that may enable the
magnitude of the kinetic SZ to be recovered more accurately and hence
allow its peculiar radial velocity to be measured to greater precision
(Haehnelt \& Tegmark
1996). Clearly, a large catalogue of radial cluster velocities
measured across the whole sky would be an invaluable resource for the
investigation of large-scale motions in the Universe.

\begin{figure}
\centerline{\epsfig{
file=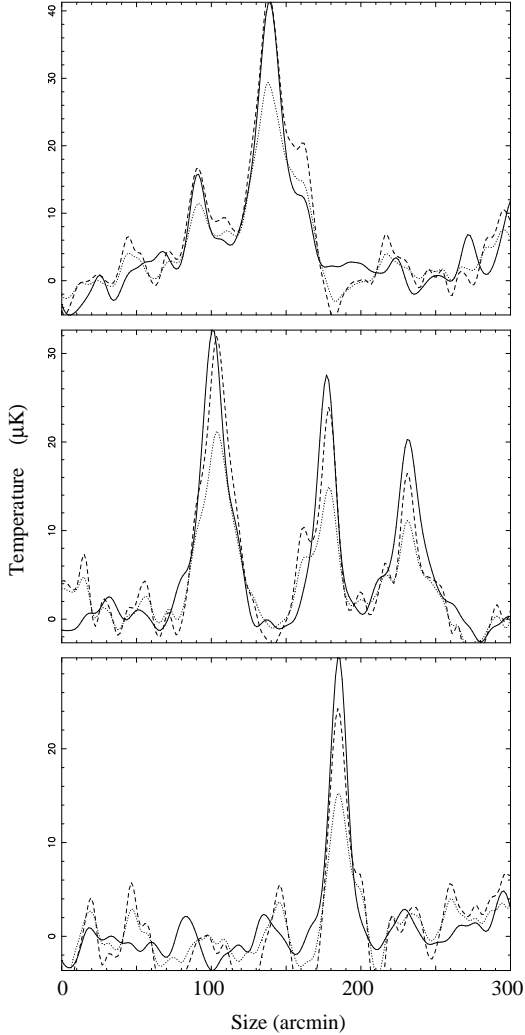,width=7cm}}
\caption{The cluster profiles of some SZ effect reconstructions
compared to the input profiles convolved with a $10\arcmin$
beam (solid line). The full MEM with full ICF information was used to
reconstruct the dashed line whereas the quadratic approximation to
this was used to reconstruct the dotted line.}
\label{fig12}
\end{figure}

Fig.~\ref{fig12} shows the MEM and WF reconstructions of the thermal
SZ profiles for a few typical clusters. These profiles are plotted as
dashed lines and dotted lines respectively and
are produced by making cuts through the reconstructed maps shown in 
Figs~\ref{fig5}(c) and \ref{fig6}(c). The reconstructions are compared
with the true cluster profiles convolved with a Gaussian
beam of FWHM $10\arcmin$, which are plotted as solid lines. 
Such a convolution is necessary in order to
make a meaningful comparison since, as we see from Fig.~\ref{fig3},
the thermal SZ effect is severely dominated by dust emission and pixel
noise in the frequency channels above 100 GHz, which have the highest
angular resolutions.  Thus the Planck Surveyor observations
contain very little information on the thermal SZ effect at
angular resolutions above about 10 arcmin.

From Fig.~\ref{fig12} we see that the MEM reconstruction of both the
peak magnitude of the SZ effect and the cluster profile are closer to
the true maps than those produced by the WF. We note that,
as expected, the WF underestimates the magnitude of the
effect and reconstructs profiles that are far less peaked.
By allowing for the bias inherent in the WF method, it is possible
to increase the heights of the main peaks in the reconstruction, but
only at the cost of increasing the overall rms residuals
significantly, since the signal-to-noise ratio for a given 
WF reconstruction is fixed.

At first sight, it 
appears that the MEM reconstructions contain several spurious
features as compared to the input profiles. This appears to have
occurred most dramatically in the top panel of the figure, on the
right-hand side of the central cluster profile. In fact, this
phenomenon illustrates the care that must taken in interpreting plots
of this type, since this feature is in fact present in the true map,
but has been smoothed out by convolving the image to 10 arcmin
resolution. The reason it is present in the MEM reconstruction is that
the effective resolution of the MEM (and WF) reconstructions can vary
across the map, depending on the level of the recovered process
compared to the other processes and the pixel noise. Thus, in some
regions, some super-resolution is possible which leads to the
reconstruction of features that are considerable smoothed by the
convolution with the 10 arcmin beam. In different regions, however,
where the other physical components happen to have high levels of
emission, or the level of pixel noise is greater, than this
super-resolution does not occur.

\subsection{Reconstructions with no ICF information}
\label{recnoicf}

Throughout subsection \ref{recicf}, the reconstructions were made assuming
full ICF information, which consists of a knowledge of the
azimuthally-averaged power spectrum of each input map, together with
cross-correlation information. In this subsection, we consider the
opposite extreme and obtain MEM and WF reconstructions assuming
virtually no ICF information. In this case we assume no
cross-correlations between components (so that the ICF matrix
$\mathbfss{L}$ is diagonal) and initially we assume the power spectrum
of each component to be constant for all measured Fourier modes and
normalised to give approximately the observed rms fluctuation in the
corresponding map.

In this case, it is no longer possible in principle to distinguish
between the primordial CMBR fluctuations and the kinetic SZ effect,
since they have the same frequency characteristics, and initially the
same power spectrum (to within a normalisation constant). Nevertheless,
we find that by attempting to reconstruct the kinetic SZ effect in this
case, the reconstructions of the other components are not 
noticeably affected. Thus, in this section, we still attempt to
reconstruct all six components. For the MEM solution
the reconstruction process is iterated, as discussed in Section
\ref{update}, but this is not possible for the WF technique since the
solution in this case tends to zero. Hence, for the WF only the
original solution is presented. For the MEM reconstruction,
the Bayesian value
of the regularising parameter $\alpha$ that satisfies (\ref{crit}) 
is found to be $\alpha = 2.9$.

\subsubsection{The reconstructed maps}

\begin{figure*}
\centerline{\epsfig{
file=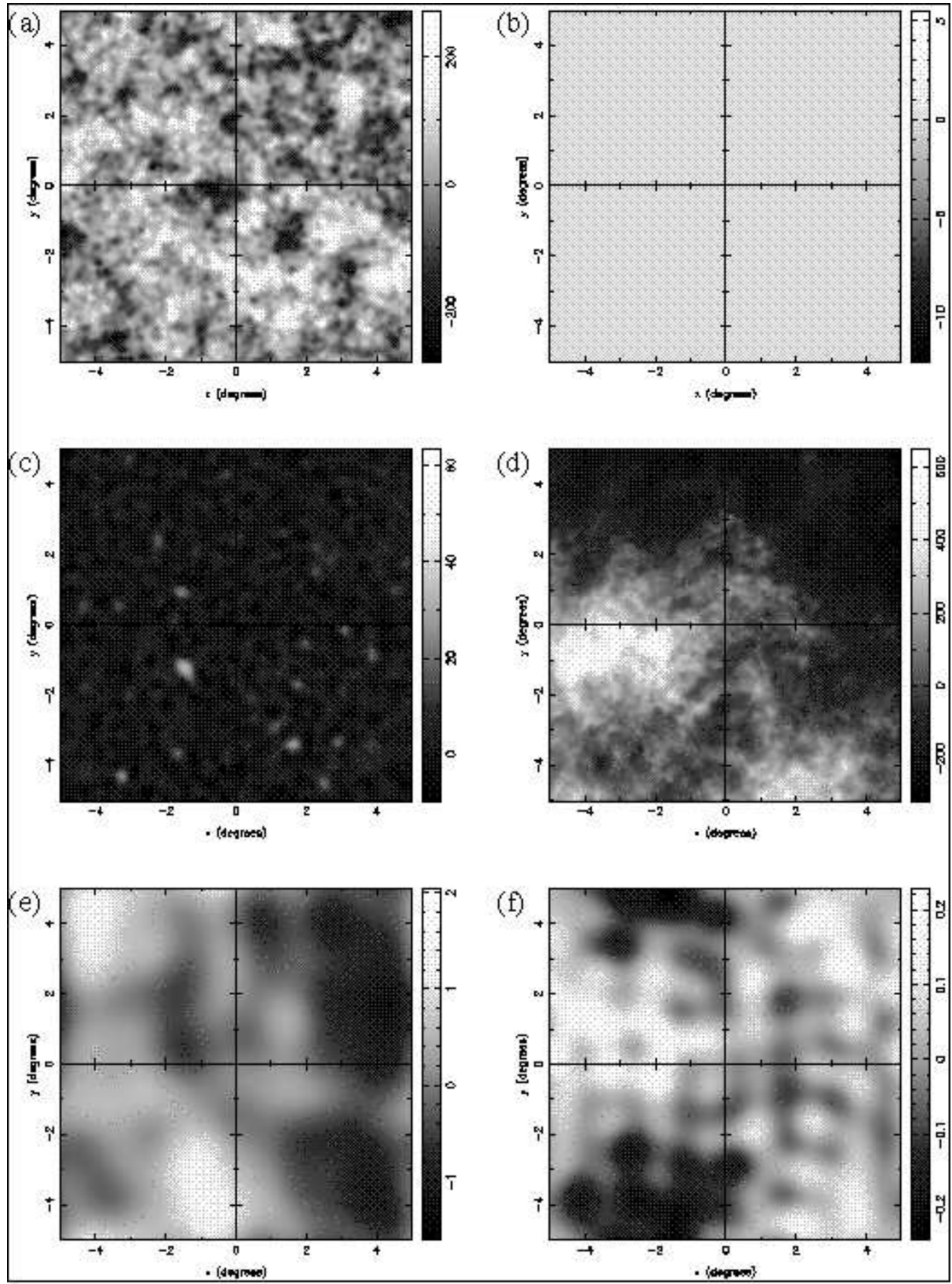,width=16cm}}
\caption{MEM reconstruction of the $10\times 10$-degree maps of
the input components shown in Fig.~\ref{fig1}, using no power
spectrum information (see text). The components are:
(a) primary CMBR fluctuations; (b) kinetic SZ effect; 
(c) thermal SZ effect; (d) Galactic dust; (e) Galactic
free-free; (f) Galactic synchrotron emission. 
Each component is plotted at 300 GHz
and has been convolved with a Gaussian beam of FWHM equal to 4.5
arcmin. The map units are equivalent thermodynamic temperature in
$\mu$K.}
\label{fig13}
\end{figure*}

\begin{figure*}
\centerline{\epsfig{
file=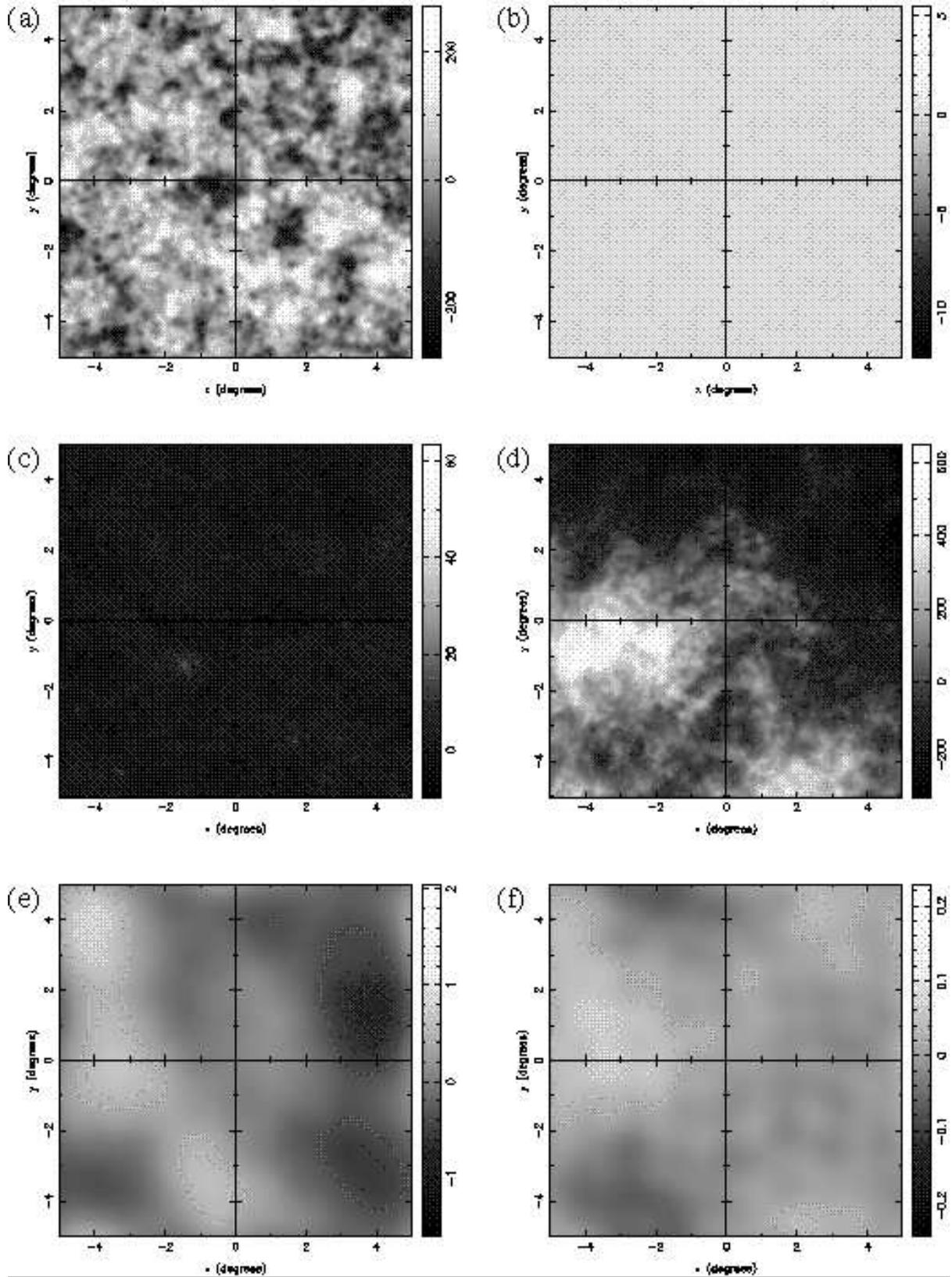,width=16cm}}
\caption{Wiener filter reconstruction of the $10\times 10$-degree maps of
the six input components shown in Fig.~\ref{fig9}, using no power
spectrum information (see text).}
\label{fig14}
\end{figure*}

Figs~\ref{fig13} and \ref{fig14} show respectively the MEM and WF
reconstructions of the six input components. Once again, for
comparison purposes, the grey scales in these figures are chosen to
coincide with those in Fig.~\ref{fig1} and
both sets of maps have beem convolved with a 4.5 arcmin FWHM Gaussian
beam. Comparing these figures with
the input maps, we see that by assuming no ICF information, the
overall quality of the reconstructed maps has been
somewhat reduced, in particular for the WF.

It is encouraging to note that both the MEM and WF reconstructions of
the CMBR, shown in panel (a) of each figure, still closely resemble
the true input map. This is also true for the reconstructions of the
dust emission shown in panel (d) of each figure.  As mentioned in
section~\ref{simobs}, it is possible, by
simple visual inspection of the data maps at each observing frequency,
to distinguish the CMBR and dust contributions quite clearly,
and so we would indeed hope that any reasonable separation algorithm
would be able to reconstruct these components with some accuracy.

The quality of both the MEM and WF reconstructions of the free-free
and synchrotron components has been significantly reduced by assuming
no ICF information. We do see, however, that both reconstructions of
the free-free component contain the main features of the input map,
but smoothed to a much lower resolution, but that MEM reconstruction
contains slightly more detail.
For the synchrotron component, shown in
panel (f), the WF algorithm has again produced a very low-level, smoothed
reconstruction of the input map, whereas the MEM reconstruction has
been more successful in recovering the brightest regions.

As expected, the quality of the MEM and WF reconstructions differs most
for the thermal SZ effect, shown in panel (c) of each figure.
Although the MEM
reconstruction is not as accurate as that obtained assuming full ICF
information, it still provides a reasonable representation of the main
features of the input map. This is certainly not true for the WF
reconstruction which contains only very low-level features at the
positions of the few largest peaks.
 

The rms of the residuals for each set of reconstructions are given
in Table~\ref{erms2}.
\begin{table}
\begin{center}
\caption{The rms residuals per 4.5 arcmin FWHM Gaussian beam
(in $\mu$K) for the MEM and WF
reconstructions shown in Figs~\ref{fig13} and \ref{fig14}, which
assume no ICF information.}
\begin{tabular}{lccc} \hline
Component & $e^{\rm MEM}_{\rm rms}$ & $e^{WF}_{\rm rms}$ 
& $e^{\rm MEM}_{\rm rms}$ (1 iter.) \\ \hline
CMBR             & 6.10 & 7.50 & 7.10 \\
Kinetic SZ       & 0.85 & 0.85 & 0.85 \\
Thermal SZ       & 4.35 & 4.61 & 4.42 \\
Dust             & 1.90 & 2.10 & 2.07 \\
Free-Free        & 0.44 & 0.50 & 0.48 \\
Synchrotron      & 0.07 & 0.08 & 0.07 \\ \hline
\end{tabular}
\end{center}
\label{erms2}
\end{table}
We see from the table that the MEM reconstruction of the CMBR has a
significantly lower rms error than the WF reconstruction and is only
marginally less accurate than that obtained assuming full ICF
information. Indeed, once again, the rms error of the MEM
reconstructions of the other components are again consistently lower
than the corresponding WF reconstructions.

As mentioned above, 
the MEM technique is iterated until the reconstructions
coverged, but this is not directly possible 
for the WF. In is therefore of some interest to investigate how much
the MEM solution is improved by this iterative process. Therefore, in
Table~\ref{erms2}, we also quote the rms residuals for the MEM
reconstruction after just one iteration. As we might expect, the
initial rms errors are slightly better than those found using the WF,
but we also see that iterating the MEM technique clearly reduces the 
residuals, most notably for the CMBR reconstruction.

\begin{figure*}
\centerline{\epsfig{
file=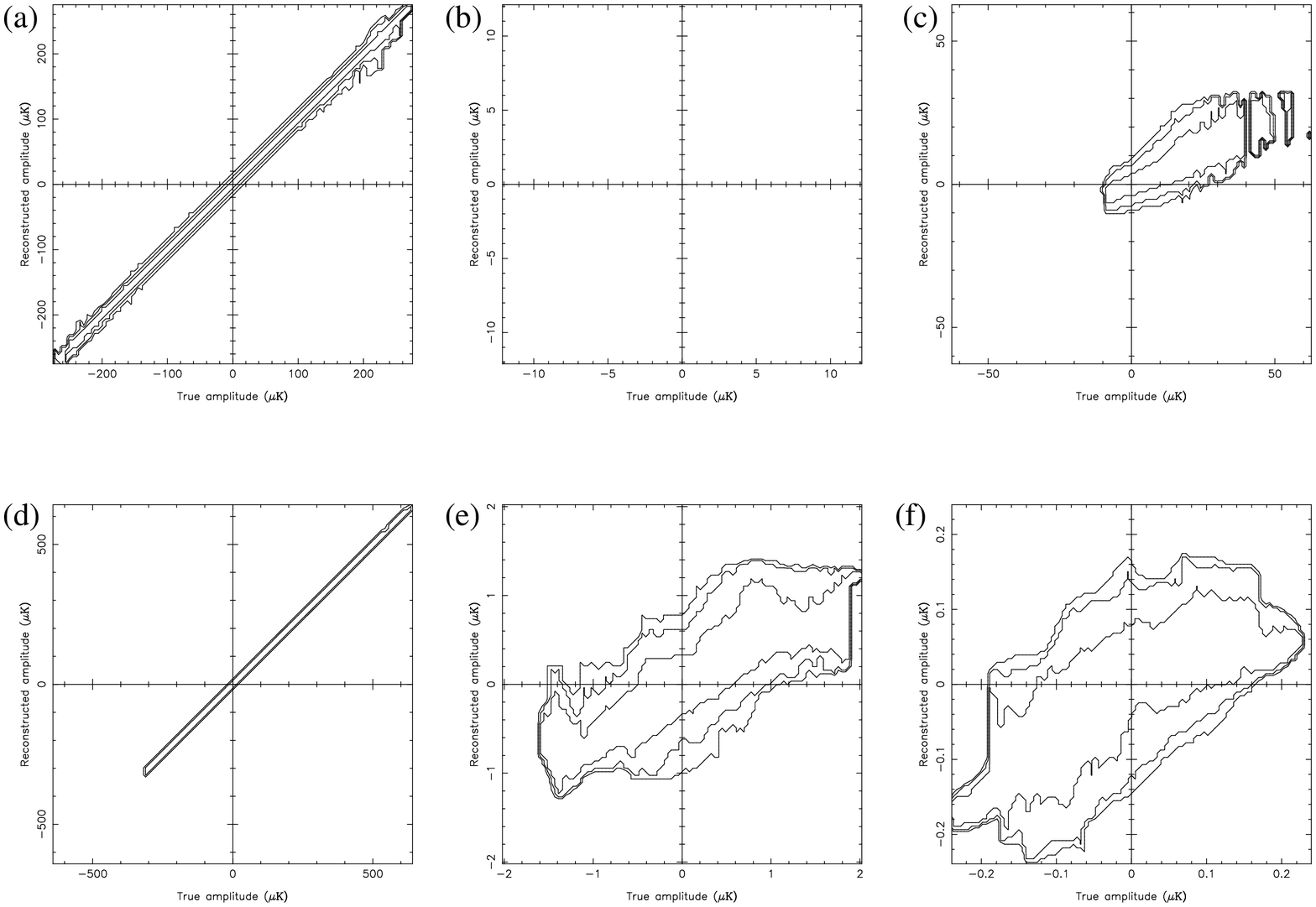,width=15.0cm}}
\caption{Comparison of the input maps with the maps reconstructed 
using the MEM algorithm with no ICF information.
The horizontal axes show the input map amplitudes 
and the vertical axes show the corresponding reconstructed amplitudes. 
The contours contain 68, 95 and 99 per cent of the pixels respectively.}
\label{fig15}
\end{figure*}
\begin{figure*}
\centerline{\epsfig{
file=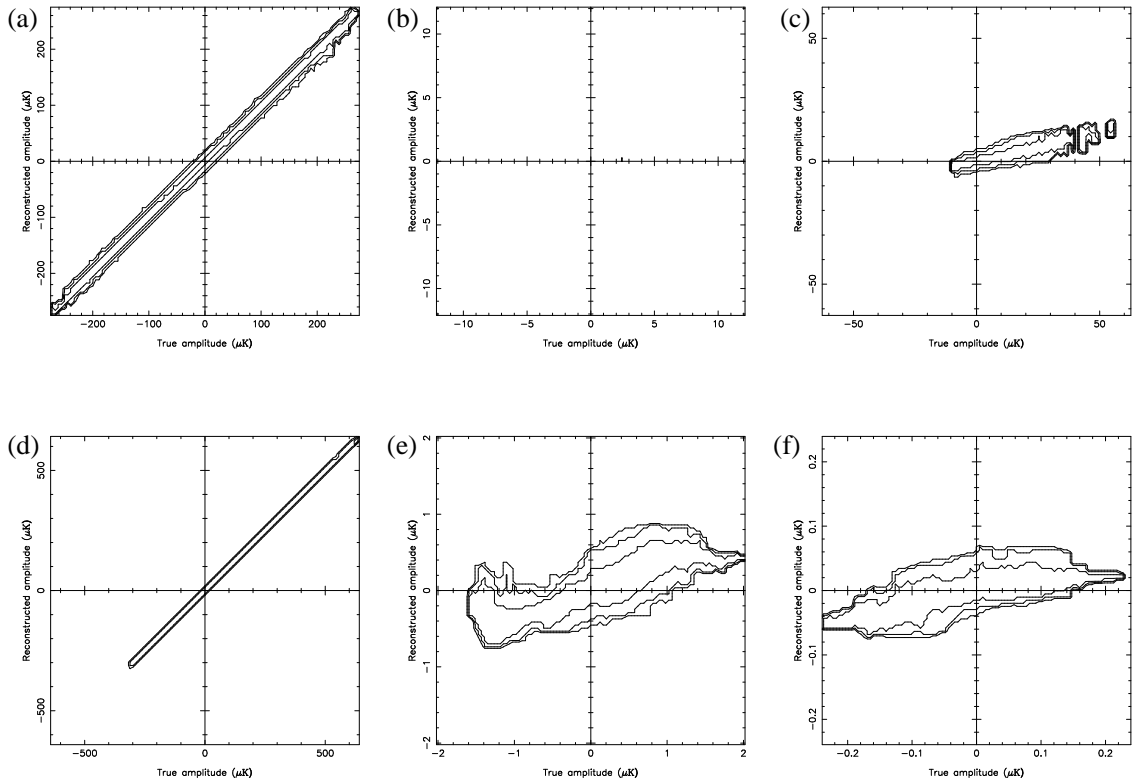,width=15.0cm}}
\caption{As for Fig.~\ref{fig15}, but for the Wiener filter
reconstruction with no ICF information.}
\label{fig16}
\end{figure*}

Figs~\ref{fig15} and \ref{fig16} show the distribution of pixel temperatures in
the MEM and WF reconstructions as compared to the pixel temperatures
in the corresponding input maps. Table~\ref{slope2} gives the gradient
of the best-fit straight line through the origin for each component.

The confidence limits for pixel temperatures in the reconstructed CMBR
maps are shown in panel (a) in each figure. We see that for most input
temperatures the confidence contours are somewhat narrower for the MEM
reconstruction than for the WF, and this is reflected in its lower
$e_{\rm rms}$ value.  At high input temperatures, however, the 95 and
99 per cent limits become slightly wider for the MEM
reconstruction. From Table~\ref{slope2} we also notice
that the best-fit straight line through the
origin has a slope of approximately 0.96 for the WF as compared to a
value of 1.0 for the MEM reconstruction. Thus in the absence of ICF 
information the WF reconstruction slightly underestimates the true 
temperature in each pixel of the CMBR map.

Panel (c) in Figs~\ref{fig15} and \ref{fig16} shows the confidence
limits for the reconstructions of the thermal SZ.  We see for the MEM
algorithm that the confidence contours are slightly wider than those in
Fig.~\ref{fig7}(c), obtained assuming full ICF information. The
best-fit straight line through the origin again has a slope
significantly smaller than unity, indicating that magnitudes of the
thermal SZ effects are underestimated, but
its slope is close to that obtained with full ICF information.
For the WF reconstruction,
however, the best-fit line now has a slope very close to zero.

The confidence contours for the WF and MEM reconstructions of the dust
component, shown in panel (d) of each figure, are again
indistinguishable and clearly show that the dust is once more the most
accurately reconstructed component.  Finally, from panels (e) and (f)
in Fig.~\ref{fig16}, we see that the slope of the confidence contours
is much closer to unity for the MEM reconstructions than for the WF
(see Table~\ref{slope2}). We note, however, that about the best-fit
line the spread of values in the WF reconstructions is smaller than
for MEM. Nevertheless, the spread of reconstructed temperatures about
the {\em true} values is still smaller for MEM, as seen by the
relative values of $e_{\rm rms}$ given in Table~\ref{erms2}.

\begin{table}
\begin{center}
\caption{The gradients of the best-fit straight line through the
origin for the comparison plots shown in Figs ~\ref{fig15} and
\ref{fig16} for the MEM and WF reconstructions, which assume no ICF information.}
\begin{tabular}{lcc} \hline
Component        &   MEM gradient    &    WF gradient     \\ \hline
CMBR             & $1.00$  &  $0.97$  \\
Kinetic SZ       & $0.00$  &  $0.00$  \\
Thermal SZ       & $0.50$  &  $0.24$  \\
Dust             & $1.00$  &  $1.00$  \\
Free-Free        & $0.60$  &  $0.22$  \\
Synchrotron      & $0.47$  &  $0.13$  \\ \hline
\end{tabular}
\end{center}
\label{slope2}
\end{table}

\subsubsection{The reconstructed power spectra}

For the reconstruction presented in this section the assumed 
power spectra of the input components were constant for all measured
Fourier modes. It is therefore of particular interest to investigate the
power spectra of the reconstructed maps in this case.

\begin{figure*}
\centerline{\epsfig{
file=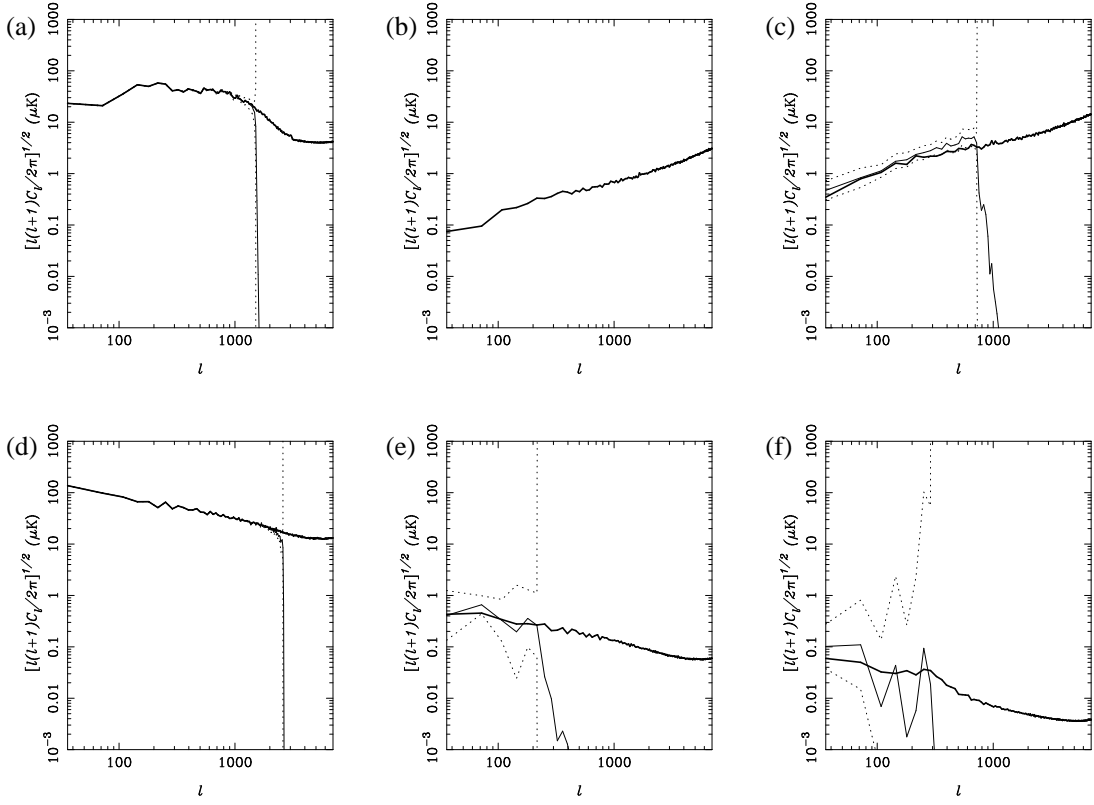,width=14.5cm}}
\caption{The power spectra of the input maps (bold line) compared to
to the power spectra of the maps reconstructed using MEM with no ICF
information (faint line). The dotted lines show one sigma confidence
limits on the reconstructed power spectra.}
\label{fig17}
\end{figure*}

\begin{figure*}
\centerline{\epsfig{
file=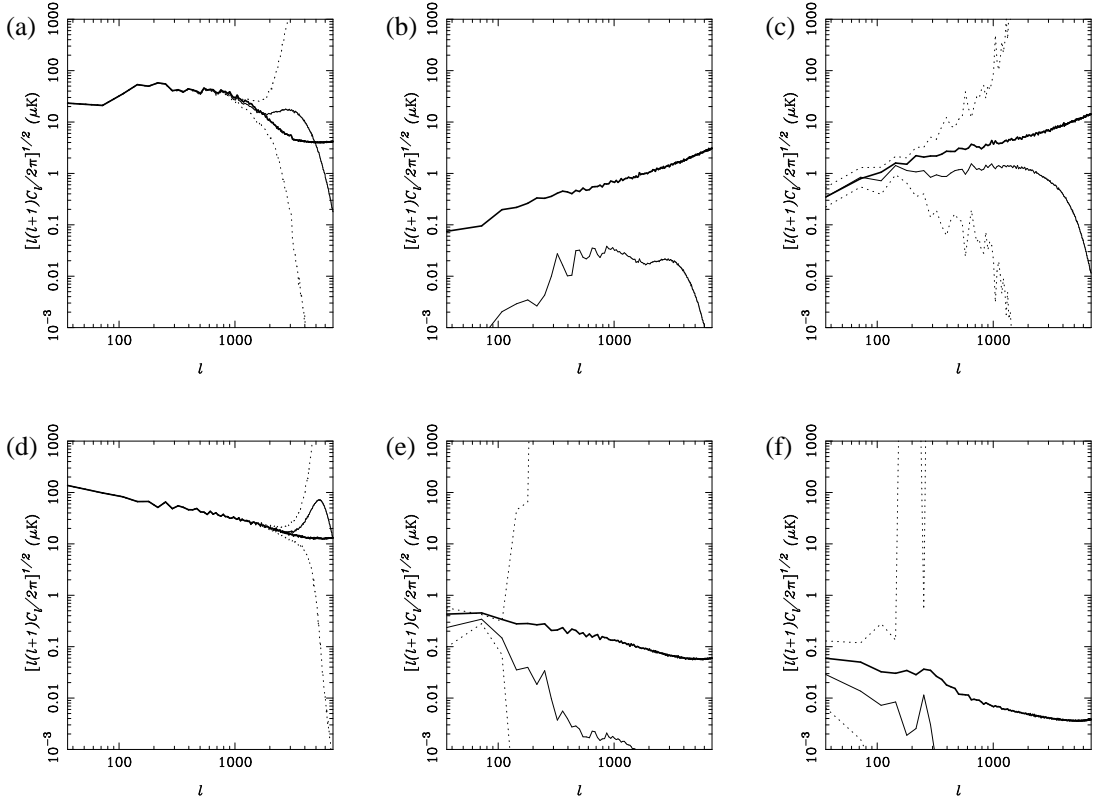,width=14.5cm}}
\caption{As for Fig~\ref{fig17}, but for the WF reconstruction with
no ICF information.}
\label{fig18}
\end{figure*}

The reconstructed power spectra are calculated in the same manner as
that outlined in subsection~\ref{recicf}, as are the errors bars. The resulting
power spectra are plotted in Figs~\ref{fig17} and \ref{fig18} for the
MEM and WF reconstructions respectively.

The power spectrum of the reconstructed CMBR maps are shown in panel
(a) of each figure, and we see that the MEM and WF techniques produce
noticeably different results. For the MEM reconstruction the power
spectrum closely follows the true spectrum out to $\ell \approx 1500$,
at which point it drops rapidly to zero. For the WF reconstruction,
however, the features in the power spectrum match those in the
true spectrum for $\ell \la 1000$, and then slightly underestimate the
true power for $\ell \approx 1000$--1500.
At higher multipoles, the power spectrum of the 
WF reconstruction contains a spurious hump, which results in an
overestimate of the true power spectrum for $\ell \approx 2000$--$5000$,
before the power spectrum finally tends to zero.

For MEM and WF reconstructions of the thermal SZ map, the
corresponding power spectra are shown in panel (c) of each figure. 
We see that the power spectrum of the MEM reconstruction is reasonably
accurate out to $\ell\approx 1000$, but does overestimate the power
slightly over this range. At higher multipoles, we again find that
the MEM power spectrum drops rapidly to zero. The power spectrum of
the WF reconstruction underestimates to true power at high multipoles
and is only reasonably accurate for $\ell \la 200$.

Panels (d), (e) and (f) in Fig.~\ref{fig17} and \ref{fig18} show the
power spectra of the MEM and WF reconstructions of the Galactic dust,
free-free and synchrotron. For the dust component, we see that the
power spectrum of the MEM component follows the true power spectrum
up to $\ell \approx 2000$, before dropping rapidly to zero. For the WF
reconstruction, however, the recovered power spectrum is accurate 
up to $\ell \approx 3000$, but then exhibits a spurious hump which
results in the overestimation of the true power at all higher
multipoles. The power spectra of the reconstructed free-free maps
are shown in panel (e) of each figure. We see that the MEM
reconstruction is accurate for $\ell \ga 100$, but then underestimates
the true power at higher multipoles, whereas the WF reconstructions
underestimates the true power at all multipoles.
For the synchrotron component, the WF reconstructions
underestimate the true power at all multipoles, whereas the MEM
solution
oscillates widely about the true power spectrum for $\ell \la 300$,
before dropping to zero.

\subsubsection{The reconstructed thermal SZ profiles}

From Figs~\ref{fig13} and \ref{fig14} we see that assuming no ICF information
leads to a substantial difference in the quality of the MEM and WF
reconstructions of the thermal SZ effect. We find that the MEM reconstruction
is only marginally less accurate than that
obtained assuming full ICF information, but the WF reconstruction is
considerably poorer in this case. 

\begin{figure}
\centerline{\epsfig{
file=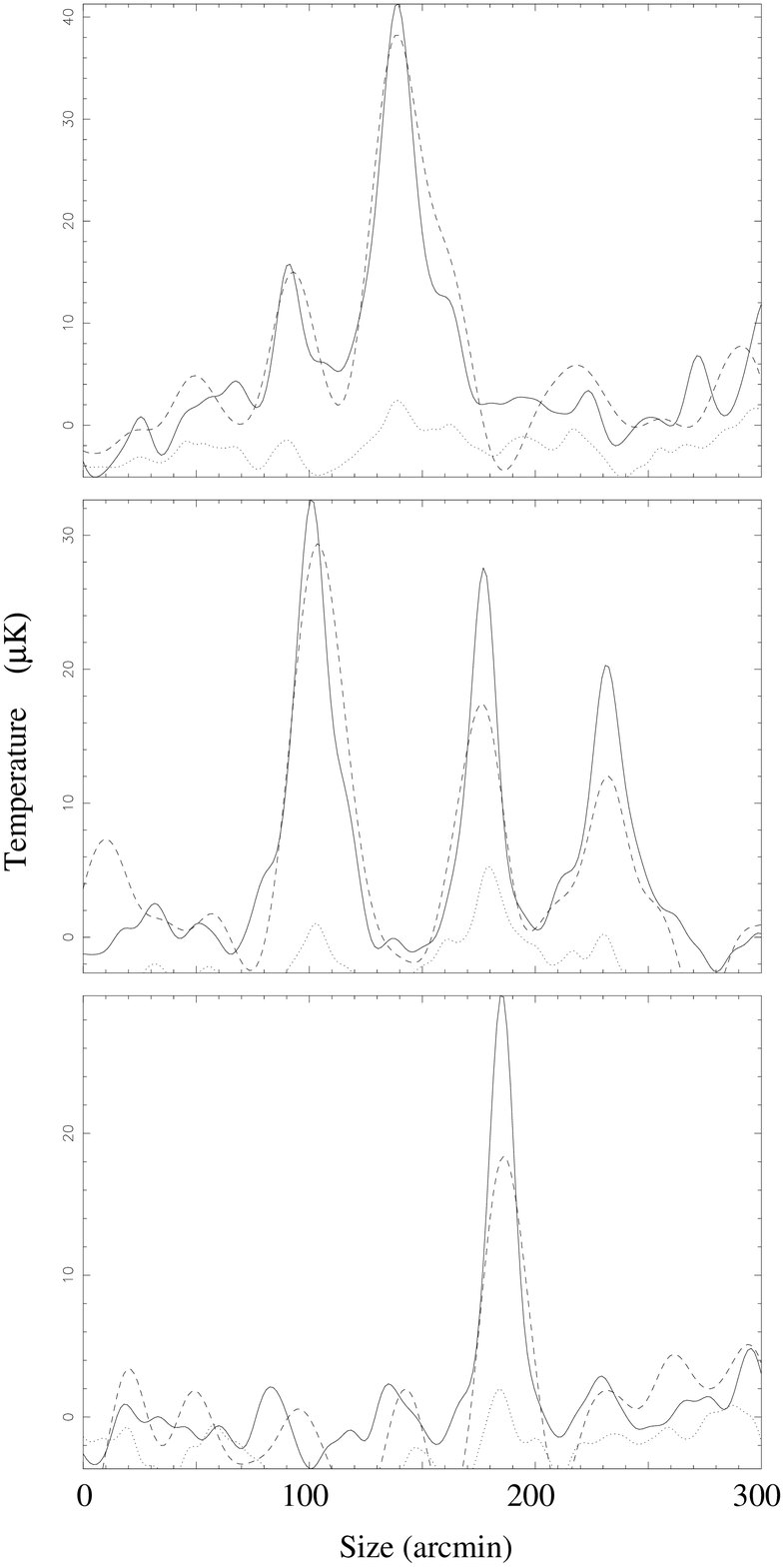,width=7cm}}
\caption{The cluster profiles of some SZ effect reconstructions
compared to the input profiles convolved with a $10\arcmin$
beam (solid line). The full MEM with full ICF information was used to
reconstruct the dashed line whereas the quadratic approximation to
this was used to reconstruct the dotted line.}
\label{fig19}
\end{figure}

Fig.~\ref{fig19} shows cuts through the MEM and WF reconstructions
that coincide with several typical clusters. The reconstructed MEM and
WF cluster profiles are plotted as dashed lines and dotted lines
respectively and are again compared with the true cluster profiles
convolved with a Gaussian beam of FWHM $10\arcmin$ (solid line). From
the figure we see that there is indeed a considerable difference
between the MEM and WF reconstructions. The cluster profiles in the
MEM reconstruction are reasonable approximations to the input
profiles, although the reconstructed peak values are slightly lower in
most cases.  For the WF reconstruction, however, the cluster profiles
are very poorly approximated indeed, with the peak value often
underestimated by an order of magnitude. Of course, this simply
reflects that it would be extremely ill-advised to use WF for determining weak
processes in the absence of the power spectrum information of
which WF is meant to take advantage.

\section{Discussion and conclusions}
\label{conc}

In this paper we adopt a Bayesian approach to the separation of
foreground components from CMBR emission for satellite observations.
In particular, we use simulated Planck Surveyor observations 
of a $10\times 10$ degree patch of sky at ten different
observing frequencies performed by Gispert \& Bouchet (1997) and Bouchet \&
Gispert (1998) . The sky emission includes contributions from  
primary CMBR fluctuations, kinetic and thermal Sunyaev-Zel'dovich
effects from clusters and dust, free-free and synchrotron emission
from the Galaxy.

We find that by assuming a suitable Gaussian prior in Bayes' theorem
for the sky emission, we recover the standard Wiener filter (WF)
approach.  Alternatively, we may assume an entropic prior, based on
information-theoretic considerations alone, from which we derive a
maximum entropy method (MEM). We apply these two methods to the 
problem of separating the different physical components of sky
emission.

The reconstructions presented in Section~\ref{results}  show that, in the
absence of severe point source contamination, the Planck Surveyor
observations enable the recovery of the CMBR fluctuations with an
absolute accuracy of about 6 $\mu$K. Moreover, depending on assumed
knowledge of the power spectra of the various components, we find that
it is possible to reconstruct the emission due to other components
with varying degrees of accuracy. In particular, the Galactic dust
emission may be reconstructed with an accuracy of about 2 $\mu$K. The
main features of Galactic free-free and synchrotron are also
reconstructed.  We find that both the magnitude and radial profile of
the thermal SZ effect may be recovered for rich clusters, but the
reconstruction of the kinetic SZ effect is only possible in clusters
which also have a large thermal SZ effect.  Given the
cluster gas profile derived from the thermal SZ effect, however, it
may be possible to recover the kinetic effect more successfully
by using optimal filtering methods tailored to individual cluster
shapes (Haehnelt \& Tegmark 1996, Aghanim et al. 1997).
We also find that the power spectra of the input components are 
well-recovered. Irrespective of the amount of prior information
assumed, we find that the CMBR power spectrum is faithfully reproduced
up to $\ell \approx 2000$, where as the 
recovered dust and thermal SZ power spectra are accurate up to
$\ell \approx 3000$ and $\ell \approx 1000$ respectively.

In nearly all cases, we find that the MEM algorithm produces equally or
more accurate reconstructed maps and power spectra of the various
components than the WF, and this is particularly true for
reconstructions of the thermal SZ effect.  This difference is most
likely a result of the assumption in the WF method that the fields to
be reconstructed are well-described by Gaussian random fields. This is
clearly not the case for the SZ effect, and other foreground
components such as Galactic dust also appear quite non-Gaussian in
nature. In the case of Galactic dust, however, the information
provided by the three highest Planck Surveyor observing frequencies
allows the WF also to recover this process very accurately.
The superiority of MEM is most apparent for processes which are both
weak and non-Gaussian. 

\subsection{Variations on standard Wiener filtering}

By assuming a Gaussian prior in Bayes' theorem, we derived the
standard form of the Wiener filter. This approach is optimal
in the sense that it is the linear filter for which the variance
of the reconstruction residuals is minimised. This is true both in 
the Fourier domain and the map domain. Nevertheless, as mentioned in
section \ref{recicf}, it is straightforward to show that this algorithm
leads to maps with power spectra that are biased compared to the true
spectra, and this leads us to consider variants of the standard
Wiener filter. 

The bias in the power spectrum of the standard WF map reconstruction
may be quantified by introducing, for each physical component, 
a {\em quality factor} $Q_p(\bmath{k})$ at each Fourier mode
$\bmath{k}$ (Bouchet et al. 1997). This factor is given by
\[
Q_p(\bmath{k}) = \sum_{\nu} W_{p\nu}(\bmath{k})R_{\nu p}(\bmath{k}),
\]
where ${\mathbfss R}(\bmath{k})$ is the response matrix of the
observations at the Fourier mode $\bmath{k}$, as defined in equation
(\ref{dataft}), and ${\mathbfss W}(\bmath{k})$ is the corresponding Wiener
matrix given in equation (\ref{wfrecon}). The quality factor varies between
unity (in the absence of noise) and zero.  If $\hat{s}_p(\bmath{k})$
is the WF estimate of the $p$th component of the signal vector at
$\bmath{k}$ and $s_p(\bmath{k}$ is the actual value, then it is
straightforward to show that 
\[
\langle |\hat{s}_p(\bmath{k})|^2 \rangle
= Q_p(\bmath{k}) \langle |s_p(\bmath{k})|^2 \rangle .
\]
Thus, in similar
way, the expectation value of the naive power spectrum estimator
defined in (\ref{psest}) is given by $\langle\hat{C}_p(k)\rangle = Q_p(k)
\langle C_p(k)\rangle$, where $Q_p(k)$ is the average of the quality
factors at each Fourier mode satisfying $|\bmath{k}|=k$; thus the
estimator in equation (\ref{psest}) is biased and should be replaced by
$\hat{C}_p(k)/Q_p(k)$. In addition, $Q_p$ may be considered as the
effective $\ell$-space window of the experiment for the process $p$. 

It is clearly unsatisfactory, however, to produce reconstructed maps
with biased power spectra and, from the above discussion, we might
consider using the matrix with elements $W_{p\nu}/Q_p^{1/2}$
to perform the reconstructions. Bouchet et al. (1997) shows that this
leads to reconstructed maps that do indeed possess unbiased power
spectra and, moreover, the method is less sensitive to the assumed
input power spectra. However, one finds in this case that the
variance of the reconstruction residuals is increase by a factor
$2(1-Q_p^{1/2})/(1-Q_p)$ compared to those obtained with the standard WF and 
so the reconstructed maps appear somewhat noisier.

Another variant of the Wiener filter technique has been proposed
by Tegmark \& Efstathiou (1996), and uses 
the matrix $W_{p\nu}/Q_p$ to perform the reconstructions. This
approach has the advantage that the reconstruction of the $p$th
physical component is independent of its assumed input power
spectrum. Nevertheless, Bouchet et al. (1997) show that the variance of the
reconstruction residuals for this technique is then increased by the
factor $(1/Q_p - 1)/(1-Q_p)$ as compared to the standard WF, which
results in even noisier reconstructed maps. 

As a final variant, Tegmark (1997) suggests the inclusion into the WF
algorithm of a parameter $\eta$ that scales the assumed input power
spectra of the components, This parameter can be included in all
of the versions of the WF discussed above and is equivalent to
assuming in Bayes' theorem a Gaussian prior of the form
\[
\Pr({\mathbfss s}) \propto 
\exp\left(-\eta{\mathbfss s}^\dagger{\mathbfss C}^{-1}{\mathbfss s}\right).
\]
In the use of this variant for the analysis of real data, 
$\eta$ is varied in order to obtain some desired
signal-to-noise ratio in the reconstructed maps by artificially
suppressing or enhancing the assumed power in the physical components
as compared to the noise. Clearly, $\eta$ plays a similar role in
the WF analysis to the parameter $\alpha$ in the MEM. Thus, by making
the appropriate changes to the calculation of the Bayesian
value of $\alpha$ in Appendix B, we may obtain an analogous expression
to (\ref{crit}) that defines a Bayesian value for $\eta$. Indeed, with
the inclusion of the parameter $\eta$, the WF method is simply the
quadratic approximation to the MEM, as discussed in Section \ref{slimit}.
However, even with the inclusion
of the $\eta$ factor, we find that the corresponding reconstructions
of non-Gaussian components are still somewhat poorer than for MEM.

\subsection{Uncertainties in spectral behaviour}

In creating the reconstructions presented in the is paper, we have
throughout assumed that the frequency dependence of all the components
are known {\em a priori}. This is a reasonable for the CMBR emission,
as well as the kinetic and thermal SZ effects, but it is unlikely to
be the case for the three Galactic components.  For real observations,
the spectral indices of the free-free and synchrotron emission will be
uncertain to within about 20 per cent. Moreover, the dust temperature
and emissivity may be known to even poorer accuracy.

If we assume for the moment that the frequency dependence of each
component is the same across the entire $10\times 10$ degree field,
then we find that reasonable uncertainties in the parameters describing
the Galactic components do not significantly affect our
reconstructions. In fact, we find that the dust temperature and
emissivity may be {\em determined} to within 1 per cent accuracy from
the data by including them as free parameters in either the MEM or WF
algorithm.  The resulting reconstructions of all the physical
components are virtually indistinguishable from those obtained by
assuming these parameters. Unfortunately, we find that it is not
possible to determine either the free-free or synchrotron spectral
index in this way. Nevertheless, if in the algorithm we assume a
spectral index for either component that is in error by within 20 per
cent, we find that the reconstructions of the remaining components are
virtually unaffected.  The resulting reconstructions of the free-free
and synchrotron components are, however, slightly poorer in this
case.

It is clear that for real observations we may not assume that the
frequency dependence of the emission in each component is the same
across the field. In this case, the method must be modified slightly, 
as discussed by Tegmark \& Efstathiou (1996) and Maisinger et al.
(1997), by the introduction of additional
channels in the reconstruction. For instance, if we assume that the 
frequency dependence of the synchrotron emission is of the form
$I \propto \nu^{-\beta}$, with $\beta=-0.7\pm 0.2$, we simply
include two synchrotron channels, one with $\beta=-0.5$ and one with
$\beta=-0.9$, or even with intermediate values, and afterwards sum
over these channels to obtain the reconstructed synchrotron
map. Alternatively, one could consider deviations from the mean
spectrum as just another template to be recovered with a modified
spectral behaviour as obtained by linearising the frequency dependence
of the intensity with theses deviations (Bouchet et al. 1996). 
 
\subsection{Future improvements and modifications}

In this paper, the simulated Planck Surveyor observations were somewhat
idealised in that it was assumed that the beam at each observing
frequency was a simple Gaussian. For the real observations,
however, it is unavoidable that the beam will in fact possess sidelobes
at some level, and care must be taken to include any such features
into the analysis, in particular if these sidelobes contain emission
from any strong sources. 

The simulated observations presented here also assume that any
striping due to the scan strategy has been removed to a sufficient
level so that it may be considered negligible. In fact, it may be
possible to use MEM to perform the destriping of the maps and the
component separation {\em simultaneously}. Indeed the simultaneous
reconstruction of a deconvolved CMBR maps and the removal
of scan baselines has already
been performed using MEM in the analysis of Tenerife data 
(Jones et al. 1997).

In terms of computational speed, however, the most important
assumption made in our simulations were that the beam at each
frequency does not change shape with position on the sky This
assumption is not unreasonable in the analysis of small patches of sky
considered here, but may be questionable for all-sky maps.  The
importance of this assumption lies in the fact that the
beam-smoothing may be written as a convolution and therefore allows us
to analyse the observations entirely in the Fourier domain, where each
mode may be considered independently. As discussed in Section \ref{methods},
this means that the analysis is reduced to a large number ($400\times
400\times 6$) of small-scale linear inversion problems and so is
computationally very fast. 

If the beam is spatially varying, however,
the beam-smoothing cannot be written as a simple convolution.  In this
case the analysis should properly be performed in the sky plane, and
requires the use of sparse matrix techniques to compute the
beam-smoothing at each point on the sky as opposed to Fast Fourier
Transforms (FFTs). In addition, the matrices involved in the linear
problem are then very large indeed, since we are attempting
simultaneously to determine $400 \times 400 \times 6$ parameters by
the minimisation of a function of corresponding dimensionality.
The large dimensionality of the problem also complicates the
inclusion of power spectrum information and the determination of
Bayesian values for $\alpha$ in the MEM algorithm and $\eta$ in the
WF. Nevertheless, the authors have investigated the use of MEM and the WF
in this case and find that reconstructions similar to those presented
here can be performed in about 12 hours of CPU on a SPARC 20
workstation. For both MEM and WF, however, the calculation of errors
cannot be performed by inverting the Hessian matrix of the posterior
probability, since this matrix has dimensions $(400 \times 400 \times
6)^2$. Although this matrix is in fact reasonable sparse, the
inversion is still not feasible. Instead, the errors on the
reconstructions must be estimated by performing several hundred
Monte-Carlo simulations for different noise realisations (see
Maisinger et al. 1997). A full discussion of the performance of 
sky-plane MEM and WF algorithms, when applied to simulated Planck
Surveyor observations, will be presented in a forthcoming paper.

Finally, perhaps the most important improvement on the simulations
and reconstructions presented here is the inclusion of 
a realistic population of point sources. Using the point source
simulations of Toffolatti et al (1998), a full investigation of
the effects on the reconstruction of the CMBR and other components
is given by Hobson et al. (in preparation).

\section*{Acknowledgements}

MPH and AWJ acknowledge Trinity Hall and King's College, Cambridge,
respectively, for support in the form of Research Fellowships. FRB
thanks R. Gispert for permission to use some of their unpublished
results.

\appendix

\section{Calculation of derivatives}
\label{appen:a}

As discussed in Section \ref{maxpost}, maximising the posterior
probability for the WF and MEM cases is equivalent to minimising 
respectively the functions $\Phi_{\rm WF}$ and $\Phi_{\rm MEM}$, which
are given by (\ref{wffunc}) and (\ref{memfunc}) as
\begin{eqnarray}
\Phi_{\rm WF}({\mathbfss h}) 
& = & \chi^2({\mathbfss h})+{\mathbfss h}^\dagger{\mathbfss h},
\label{wffuncap} \\
\Phi_{\rm MEM}({\mathbfss h}) 
& = & \chi^2({\mathbfss h})-
\alpha S_c({\mathbfss h},{\mathbfss m}_u,{\mathbfss m}_v).
\label{memfuncap}
\end{eqnarray}
From (\ref{chi2def2}), in each case the standard $\chi^2$ misfit
statistic may be written in terms of the hidden vector 
$\mathbfss{h}=\mathbfss{L}^{-1}\mathbfss{s}$ as
\begin{equation}
\chi^2({\mathbfss h}) = 
({\mathbfss d}-{\mathbfss R}{\mathbfss L}{\mathbfss h})^\dagger 
{\mathbfss N}^{-1} ({\mathbfss d}-{\mathbfss R}{\mathbfss L}{\mathbfss h}).
\label{chi2def2ap}
\end{equation}
The cross entropy $S_c({\mathbfss h},{\mathbfss m}_u,{\mathbfss m}_v)$
for this complex image is given by (\ref{entdef}) and (\ref{totent}).

Since $\mathbfss{h}$ is a complex vector we may consider 
$\Phi_{\rm WF}$ and $\Phi_{\rm MEM}$ to be functions
of the real and imaginary parts of the elements of $\mathbfss{h}$.
Alternatively, we may consider these functions to depend
upon the complex elements of $\mathbfss{h}$, together with their complex
conjugates. While it is clear that the former approach is required in
order to use standard numerical minimisers, a simpler mathematical
derivation is provided by adopting the latter approach. In any case,
derivatives with respect to the real and imaginary parts of
$\mathbfss{h}$ may be easily found using the relations
\begin{eqnarray*}
\nabla_{\Re({\mathbfss h})} & \equiv & 
\nabla_{{\mathbfss h}} + \nabla_{{\mathbfss h}^*},\\
\nabla_{\Im({\mathbfss h})} & \equiv & 
{\rm i}\left(\nabla_{{\mathbfss h}} - \nabla_{{\mathbfss h}^*}\right).
\end{eqnarray*}

Differentiating (\ref{chi2def2ap}) with respect to $\mathbfss{h}$ and
$\mathbfss{h}^*$, we find the gradient of $\chi^2$ is given by
\begin{equation}
\nabla_{{\mathbfss h}^*}\chi^2 = \left[\nabla_{\mathbfss h}\chi^2\right]^*
=-{\mathbfss L}^{\rm T}{\mathbfss R}^\dagger{\mathbfss N}^{-1}
({\mathbfss d}-{\mathbfss RLh}),
\label{chi2grad}
\end{equation}
and upon differentiating once more we find the Hessian (curvature) 
matrix of $\chi^2$ has the form
\begin{equation}
\nabla_{\mathbfss h}\nabla_{{\mathbfss h}^*}\chi^2
=
{\mathbfss L}^{\rm T}{\mathbfss R}^\dagger{\mathbfss N}^{-1}
{\mathbfss R}{\mathbfss L}.
\label{chi2curv}
\end{equation}

Using (\ref{chi2grad}) and (\ref{chi2curv}), the gradient
of $\Phi_{\rm WF}$ in (\ref{wffuncap}) is simply given by
\begin{equation}
\nabla_{{\mathbfss h}^*}\Phi_{\rm WF} 
= \left[\nabla_{\mathbfss h}\Phi_{\rm WF}\right]^*
= -{\mathbfss L}^{\rm T}{\mathbfss R}^\dagger{\mathbfss N}^{-1}
({\mathbfss d}-{\mathbfss RLh})+{\mathbfss h},\label{wfgrad}
\end{equation}
and its Hessian matrix reads
\begin{equation}
{\mathbfss H}_{\rm WF} 
= \nabla_{\mathbfss h}\nabla_{{\mathbfss h}^*}\Phi_{\rm WF}
= {\mathbfss L}^{\rm T}{\mathbfss R}^\dagger{\mathbfss N}^{-1}
{\mathbfss R}{\mathbfss L}+{\mathbfss I},\label{wfcurv}
\label{wfhess}
\end{equation}
where $\mathbfss{I}$ is the unit matrix of
appropriate dimensions. We note that the Hessian matrix for $\Phi_{\rm
WF}$ is independent of $\mathbfss{h}$. 

By setting the right-hand side of (\ref{wfgrad}) equal to zero, and
remembering that $\mathbfss{s}=\mathbfss{Lh}$ and ${\mathbfss C} =
{\mathbfss L}{\mathbfss L}^{\rm T}$, it is straightforward to obtain
the linear relation (\ref{wfrecon}) for the WF solution.  Moreover,
from (\ref{errgen}), the error covariance matrix for the reconstructed
signal vector is given by ${\mathbfss E} = {\mathbfss
LH}^{-1}{\mathbfss L}^{\rm T}$, and using (\ref{wfcurv}) this is
easily shown to be identical to the result (\ref{wferrors}).

In a similar way, we may calculate the derivatives of $\Phi_{\rm MEM}$
defined in (\ref{memfuncap}). Unfortunately, the form of the cross
entropy given in (\ref{entdef}) precludes us from writing its gradient or
curvature as a simple matrix multiplication, and we must instead
express them in component form. From (\ref{entdef}) and (\ref{totent}), we find
the components of the gradient vector of the cross entropy are given
by
\begin{eqnarray}
\pd{S_c}{h_j} 
& = & \left(\pd{S_c}{h^*_j}\right)^* \nonumber \\
& = & -{\textstyle\frac{1}{2}}\ln
\left[\frac{\Re(\psi_j +h_j)}{2\Re({m_u}_j)}\right]
-{\textstyle\frac{1}{2}}{\rm i}\ln
\left[\frac{\Im(\psi_j+h_j)}{2\Im({m_u}_j)}\right],
\label{entgrad}
\end{eqnarray}
where 
$\Re(\psi_j) = [\Re(h_j)+4\Re({m_u}_j)\Re({m_v}_j)]^{1/2}$
and a similar expression exists for $\Im(\psi_j)$.
Differentiating once more we find the components of the 
Hessian of the cross entropy to be given by
\begin{equation}
\spd{S_c}{h_j}{h^*_{k}}=
-{\textstyle\frac{1}{4}}
\left[\frac{1}{\Re(\psi_j)}+\frac{1}{\Im(\psi_j)}\right]
\qquad \mbox{if $j=k$}
\label{entcurv}
\end{equation}
and equals zero otherwise.
We note that these components may be used to define the (diagonal) 
metric on the space of images, which is given simply by
by ${\mathbfss G}({\mathbfss h}) 
= -\nabla_{\mathbfss h}\nabla_{{\mathbfss h}^*}S_c$ (Skilling 1989;
Hobson \& Lasenby 1998). 

Using (\ref{entgrad}) and (\ref{entcurv}) the
gradient and Hessian of $\Phi_{\rm MEM}$ are then easily calculated.
In particular, we find that the Hessian matrix is given by
\begin{equation}
{\mathbfss H}_{\rm MEM}
= \nabla_{\mathbfss h}\nabla_{{\mathbfss h}^*}(\chi^2
-\alpha S_c) = 
{\mathbfss L}^{\rm T}{\mathbfss R}^\dagger{\mathbfss N}^{-1}
{\mathbfss R}{\mathbfss L} + \alpha {\mathbfss G}
\label{memcurv}
\end{equation}
where $\mathbfss{G}$ is the image space metric and we have used the 
expression for the curvature of $\chi^2$ given
in (\ref{chi2curv}). In contrast to (\ref{wfhess}), we see that the 
Hessian matrix of $\Phi_{\rm MEM}$ depends on $\mathbfss{h}$
through the metric $\mathbfss{G}$.

\section{Bayesian value for $\alpha$}
\label{appen:b}

A Bayesian value for $\alpha$ may be found simply by treating it as
another parameter in our hypothesis space. This procedure is outlined
for the case of real images in Skilling (1989) and
Gull \& Skilling (1990), and we modify their treatment here in order to
accommodate complex images $\mathbfss{h}$.

After including $\alpha$ into our hypothesis space, the full joint
probability distribution can be expanded as
\begin{eqnarray}
\Pr({\mathbfss h,d},\alpha) 
& = & 
\Pr(\alpha)\Pr({\mathbfss h}|\alpha)
\Pr({\mathbfss d}|{\mathbfss h},\alpha) \nonumber \\
& = &
\Pr(\alpha)\Pr({\mathbfss h}|\alpha)
\Pr({\mathbfss d}|{\mathbfss h}) \label{fjoint}
\end{eqnarray}
where in the last factor we can drop the conditioning on $\alpha$ since
it is $\mathbfss{h}$ alone that induces the data $\mathbfss{d}$. We
then recognise this as the 
likelihood. Furthermore, the second factor
$\Pr({\mathbfss h}|\alpha)$ can be identified as the entropic prior
and so (\ref{fjoint}) becomes
\begin{eqnarray}
\Pr({\mathbfss h,d},\alpha) 
& = & \Pr(\alpha)\frac{e^{\alpha S_c({\mathbfss h})}}{Z_S(\alpha)}
\frac{e^{-\chi^2({\mathbfss h})}}{Z_L} \nonumber \\
& = & \Pr(\alpha)
\frac{e^{\alpha S_c({\mathbfss h})-\chi^2({\mathbfss h})}}{Z_S(\alpha)Z_L},
\label{fjoint2}
\end{eqnarray}
where $Z_S(\alpha)$ and $Z_L$ are respectively the normalisation
constants for the entropic prior and the likelihood such that the
total probability density function in each case integrates to unity.
For convenience we have dropped the explicit dependence of the
cross entropy $S_c$ on the models ${\mathbfss m}_u$ and ${\mathbfss m}_v$.

Since we have assumed the instrumental noise on the data to be 
Gaussian, the likelihood function is also Gaussian
and so the normalisation factor $Z_L$ is easily found. Evaluating the
appropriate Gaussian integral gives
\[
Z_L = \pi^{n_f} \det{N}
\]
where $n_f$ is the dimension of the complex data vector 
${\mathbfss d}$ and is equal to the number of observing frequencies
that make up the Planck Surveyor data set; $\det{N}$ is the
determinant of the noise covariance matrix defined in (\ref{ncovdef}).

The normalisation factor $Z_S(\alpha)$ for the entropic prior is more
difficult to calculate since this prior is not Gaussian in
shape. Nevertheless, we find that a reasonable approximation to
$Z_S(\alpha)$ for all $\alpha$ may be obtained by making a Gaussian
approximation to the prior at its maximum, which occurs at
${\mathbfss h}_m = {\mathbfss m}_u-{\mathbfss m}_v$.
As discussed in Appendix A, the Hessian matrix of the entropy 
at this point is given by 
$\nabla_{\mathbfss h}\nabla_{{\mathbfss h}^*}S_c=-{\mathbfss G}$, where
${\mathbfss G}$ is the metric on image space evaluated at the maximum
of the prior ${\mathbfss h}_m$; the metric matrix is real and diagonal.
Remembering that $S_c({\mathbfss h}_m)=0$ and 
using the Gaussian approximation,  $Z_S(\alpha)$ is then given by
\begin{eqnarray}
Z_S(\alpha) & = & 
\int_\infty e^{\alpha S_c({\mathbfss h})}
\,\det{G}\,{\rm d}^{n_c}{\mathbfss h}\,{\rm d}^{n_c}{\mathbfss h}^*
\nonumber \\
& \approx &
\int_\infty e^{-\alpha
({\mathbfss h}-{\mathbfss h}_m)^\dagger
{\mathbfss G}({\mathbfss h}-{\mathbfss h}_m)}
\,\det{G}\,{\rm d}^{n_c}{\mathbfss h}\,{\rm d}^{n_c}{\mathbfss h}^*
\nonumber \\
& \approx &
\pi^{n_c}|\alpha{\mathbfss I}|^{-1} = (\pi/\alpha)^{n_c},
\label{zadef}
\end{eqnarray}
where $n_c$ is the dimension of the complex (hidden) image vector 
${\mathbfss h}$ and is equal to the number of physical components
present in the simulations.

Now, returning to (\ref{fjoint2}), 
in order to investigate more closely the role of
$\alpha$, we begin by considering
the joint probability distribution $\Pr({\mathbfss d},\alpha)$, which
may be obtained by integrating out ${\mathbfss h}$ in (\ref{fjoint2}):
\begin{eqnarray}
\Pr({\mathbfss d},\alpha) 
& = & \int_\infty \Pr({\mathbfss h,d},\alpha) 
\,\det{G}\,{\rm d}^{n_c}{\mathbfss h}\,{\rm d}^{n_c}{\mathbfss h}^*
\nonumber \\
& = & \frac{\Pr(\alpha)}{Z_S(\alpha)Z_L}
\int_\infty 
e^{\alpha S_c({\mathbfss h})-\chi^2({\mathbfss h})}
\,\det{G}\,{\rm d}^{n_c}{\mathbfss h}\,{\rm d}^{n_c}{\mathbfss h}^*
\nonumber \\
& \equiv & \Pr(\alpha)\frac{Z_{\Phi}(\alpha)}{Z_S(\alpha)Z_L}
\label{zphidef}
\end{eqnarray}
where we have defined the normalisation integral $Z_\Phi(\alpha)$.
In order to calculate $Z_\Phi(\alpha)$, we follow a similar
approach to that use to calculate $Z_S(\alpha)$ and make a Gaussian
approximation to 
$\exp[\alpha S_c({\mathbfss h})-\chi^2({\mathbfss h})]$
about its maximum at 
$\hat{\mathbfss h}$. The required Hessian matrix ${\mathbfss H}_{\rm MEM}$
is given by (\ref{memcurv}) evaluated at $\hat{\mathbfss h}$. 
Let us, however, define a new matrix ${\mathbfss M}$ that is given by
\begin{equation}
{\mathbfss M} 
\equiv {\mathbfss G}^{-1/2}{\mathbfss H}_{\rm MEM}{\mathbfss G}^{-1/2}
={\mathbfss G}^{-1/2}
{\mathbfss L}^{\rm T}{\mathbfss R}^\dagger{\mathbfss N}^{-1}
{\mathbfss R}{\mathbfss L}{\mathbfss G}^{-1/2} + 
\alpha {\mathbfss I}.
\label{mmatdef}
\end{equation}
The integral $Z_\Phi(\alpha)$ is then approximated by
\begin{eqnarray}
Z_\Phi(\alpha) 
& \approx &
e^{\alpha S_c(\hat{\mathbfss h})-\chi^2(\hat{\mathbfss h})}
\int_\infty e^{-
({\mathbfss h}-\hat{\mathbfss h})^\dagger
{\mathbfss H}_{\rm MEM}({\mathbfss h}-\hat{\mathbfss h})}
\,\det{G}\,{\rm d}^{n_c}{\mathbfss h}\,{\rm d}^{n_c}{\mathbfss h}^*
\nonumber \\
& \approx &
e^{\alpha S_c(\hat{\mathbfss h})-\chi^2(\hat{\mathbfss h})}
\int_\infty e^{-
({\mathbfss h}-\hat{\mathbfss h})^\dagger
{\mathbfss G}^{1/2}{\mathbfss M}{\mathbfss G}^{1/2}
({\mathbfss h}-\hat{\mathbfss h})}
\,\det{G}\,{\rm d}^{n_c}{\mathbfss h}\,{\rm d}^{n_c}{\mathbfss h}^*
\nonumber \\
& \approx & 
e^{\alpha S_c(\hat{\mathbfss h})-\chi^2(\hat{\mathbfss h})}
\pi^{n_c} \det{M}^{-1}.
\label{zphival}
\end{eqnarray}

Thus, substituting into (\ref{zphidef})
the expressions for $Z_S(\alpha)$ and $Z_\Phi(\alpha)$ given
by (\ref{zadef}) and (\ref{zphival}) respectively, we find that
in the Gaussian approximation the joint
probability distribution $\Pr({\mathbfss d},\alpha)$ has the form
\begin{eqnarray*}
\Pr({\mathbfss d},\alpha) & = & \Pr(\alpha) \Pr({\mathbfss d}|\alpha) \\
& \approx & \Pr(\alpha)Z_L^{-1}
e^{\alpha S_c(\hat{\mathbfss h})-\chi^2(\hat{\mathbfss h})} 
\alpha^{n_c}\det{M}^{-1}.
\end{eqnarray*}

Now, in order to obtain a Bayesian estimate for $\alpha$, we should
choose an appropriate form for the prior $\Pr(\alpha)$. Nevertheless,
for realistically large data sets, the distribution 
$\Pr({\mathbfss d}|\alpha)$ is so strongly peaked that it overwhelms
any reasonable prior on $\alpha$, and so we assign the Bayesian value
$\hat{\alpha}$ of the regularisation constant to be that which
maximises $\Pr({\mathbfss d}|\alpha)$. Taking logarithms we obtain
\[
\ln\Pr({\mathbfss d}|\alpha)=\mbox{constant}
+\alpha S_c(\hat{\mathbfss h})-\chi^2(\hat{\mathbfss h})
+ n_c\ln\alpha - \ln\det{M}.
\]
Differentiating with respect to $\alpha$, and noting that the 
$\hat{\mathbfss h}$-derivatives cancel, we find
\begin{equation}
\nd{}{\alpha}\ln\Pr({\mathbfss d}|\alpha)=
S_c(\hat{\mathbfss h}) + \frac{n_c}{\alpha} 
- {\rm Tr}\left({\mathbfss M}^{-1}\nd{{\mathbfss M}}{\alpha}\right),
\label{crit2}
\end{equation}
where we have used the identity
\[
\nd{}{\alpha}\ln\det{M} \equiv 
{\rm Tr}\left({\mathbfss M}^{-1}\nd{{\mathbfss M}}{\alpha}\right),
\]
which is valid for any non-singular matrix $\mathbfss{M}(\alpha)$.
From (\ref{mmatdef}), however, we see that 
${\rm d}{\mathbfss M}/{\rm d}\alpha =
{\mathbfss I}$. Substituting this relation into (\ref{crit2}) and equating
to the result to zero, we find that in order to maximise 
$\Pr({\mathbfss d}|\alpha)$, the parameter $\alpha$ must satisfy
\begin{equation}
-\alpha S_c(\hat{\mathbfss h}) = n_c 
-\alpha {\rm Tr}({\mathbfss M}^{-1}).
\label{critderiv}
\end{equation}

\section{Singular value decomposition in a Bayesian context}
\label{appen:c}

As outlined by Bouchet et al. (1997) and Bouchet \& Gispert (1998), a
straightforward initial approach to the component separation problem
is to perform a singular value decomposition (SVD) at each Fourier
mode separately. A full description of
the SVD technique is given by Press et al. (1994). Generalising their
discussion slightly to include complex matrices, the SVD of the 
$n_f\times n_c$ response matrix ${\mathbfss R}$ is given by
\begin{equation}
{\mathbfss R} = {\mathbfss U}{\mathbfss W}{\mathbfss V}^\dagger,
\label{svddef}
\end{equation}
where ${\mathbfss U}$ and ${\mathbfss V}$ are unitary matrices with
dimensions $n_f\times n_c$ and $n_c\times n_c$ respectively, and 
${\mathbfss W}$ is a $n_c\times n_c$ diagonal matrix.

From (\ref{dataft2}), at each Fourier mode, we have 
$\mathbfss{d} = \mathbfss{R}\mathbfss{s}+\bmath{\epsilon}$, and the
SVD estimator of the signal vector is given by
\begin{equation}
\hat{{\mathbfss s}} = 
{\mathbfss V}{\mathbfss W}^{-1}{\mathbfss U}^\dagger{\mathbfss d}.
\label{svdrecon}
\end{equation}
It is straightforward to show that this estimator minimises
the residual $|{\mathbfss d}-{\mathbfss Rs}|$ (Press et al. 1994).
Thus, from (\ref{chi2def}), we see that the SVD solution minimises
$\chi^2({\mathbfss s})$ provided the noise covariance matrix
${\mathbfss N}$ is equal to the identity matrix. Therefore, in the
context of Bayes' theorem (\ref{bayes}), the SVD solution is equivalent
to assuming a uniform prior and {\em independent} Gaussian noise
with {\em unit variance}.

We can make the connection between the SVD and modified minimum chi-squared
solutions more explicit by rewriting the SVD solution solely in terms of the
response matrix ${\mathbfss R}$. Using the
unitary properties of the matrices ${\mathbfss U}$ and ${\mathbfss
V}$, it is easy to show that the SVD solution (\ref{svdrecon}) can be
rewritten as
\begin{equation}
\hat{{\mathbfss s}} = 
\left({\mathbfss R}^\dagger{\mathbfss R}\right)^{-1}{\mathbfss R}^\dagger
{\mathbfss d}.
\label{svdrecon2}
\end{equation}

Alternatively, we find from (\ref{chi2grad}) 
that the gradient of $\chi^2({\mathbfss s})$ with
respect to ${\mathbfss s}$ is given by
\begin{equation}
\nabla_{{\mathbfss s}^*}\chi^2 = \left[\nabla_{\mathbfss s}\chi^2\right]^*
=-{\mathbfss R}^\dagger{\mathbfss N}^{-1}
({\mathbfss d}-{\mathbfss Rs}),
\label{chi2sgrad}
\end{equation}
Equating this expression for the gradient to zero, we quickly obtain
the minimum chi-squared estimator
\begin{equation}
\hat{{\mathbfss s}} = 
\left({\mathbfss R}^\dagger{\mathbfss N}^{-1}
{\mathbfss R}\right)^{-1}{\mathbfss R}^\dagger{\mathbfss N}^{-1}
{\mathbfss d},
\label{minchisol}
\end{equation}
which, on setting ${\mathbfss N}$ equal to the identity matrix, is
identical to the SVD solution (\ref{svdrecon2}).

\bsp  
\label{lastpage}
\end{document}